\documentclass[aps,prb,reprint,groupedaddress,footinbib,superscriptaddress]{revtex4-1}
\usepackage{amsmath}
\usepackage{graphicx}
\usepackage{hyperref}
\usepackage{bbold}
\usepackage{color}
\usepackage[normalem]{ulem}

\begin{document}
	\title{A Unified Description of Spin Transport, Weak Antilocalization and Triplet Superconductivity in Systems with Spin-Orbit Coupling}
\author{Stefan Ili\'{c}}
\affiliation{Centro de F\'{i}sica de Materiales (CFM-MPC), Centro Mixto CSIC-UPV/EHU, 20018 Donostia-San Sebasti\'{a}n, Spain} 

\author{Ilya V. Tokatly}
\affiliation{Nano-Bio Spectroscopy Group, Departamento de Pol\'imeros y Materiales Avanzados: F\'isica, Qu\'imica y Tecnolog\'ia, Universidad del Pa\'is Vasco (UPV/EHU), 20018 Donostia-San Sebasti\'{a}n, Spain} 
\affiliation{IKERBASQUE, Basque Foundation for Science, 48011 Bilbao, Spain}
\affiliation{Donostia International Physics Center (DIPC), 20018 Donostia-San Sebasti\'{a}n, Spain}

\author{F. Sebasti\'an Bergeret}
\affiliation{Centro de F\'{i}sica de Materiales (CFM-MPC), Centro Mixto CSIC-UPV/EHU, 20018 Donostia-San Sebasti\'{a}n, Spain}
\affiliation{Donostia International Physics Center (DIPC), 20018 Donostia-San Sebasti\'{a}n, Spain}
	
	\begin{abstract}
 The Eilenberger equation is a standard tool in the description of superconductors with an arbitrary degree of disorder. It can be generalized to systems with linear-in-momentum  spin-orbit coupling (SOC), by  exploiting the analogy of SOC with a non-abelian background field. Such field mixes singlet and triplet components and  yields the rich physics of magnetoelectric phenomena. 
In this work we show that   the application of this equation extends further,  beyond superconductivity. In the normal state, the linearized Eilenberger equation  describes the coupled spin-charge dynamics. Moreover, its resolvent corresponds to the so called Cooperons, and can be used to calculate the weak localization corrections. 
Specifically, we show how to solve  {this equation for any source term} and provide a closed-form solution for the case of Rashba SOC. We use this solution to address several problems of interest for  spintronics and superconductivity. Firstly, we study spin injection from  ferromagnetic electrodes in the normal state, and describe the spatial evolution of spin density in the sample, and the complete crossover from the diffusive to the ballistic limit. Secondly, we address the so-called superconducting Edelstein effect, and generalize the previously known results  to arbitrary disorder. Thirdly, we study weak localization correction beyond the diffusive limit, which can be a valuable tool in experimental characterization of materials with very strong SOC. We also address the so-called pure gauge case where the persistent spin helices form.  Our work establishes the linearized Eilenberger equation as a powerful and a very versatile method for the study of materials with spin-orbit coupling, which often provides a simpler and more intuitive picture compared to alternative methods.     
	\end{abstract}

	\maketitle
	\section{Introduction\label{sec1}}

Materials and nanostructures with spin-orbit coupling (SOC) are a subject of intensive research because of their potential for application in spintronics \cite{vzutic2004spintronics}. Coupling of spin and orbital degrees of freedom leads to various magnetolectric phenomena, which allow to achieve a spin response by  applying electric fields, and vice-versa. Most well known examples of such effects are the spin Hall effect \cite{sinova2015spin, mishchenko2004spin}, spin-galvanic or inverse Edelstein effect (SGE/IEE) \cite{ivchenko1989photocurrent,ivchenko1990current}, and inverse spin-galvanic or Edelstein effect (ISGE/EE)\cite{aronov1989nuclear, edelstein1990spin}.

SOC also has important consequences in the superconducting state, particularly in non-centrosymmetric superconductors \cite{yip2014noncentrosymmetric,smidman2017superconductivity,bauer2012non}. Namely, SOC  induces a mixing between  singlet and triplet correlations\cite{gor2001superconducting}.  Breaking time-reversal symmetry in these superconductors may lead to the formation of modulated helical phases, \cite{edel1989characteristics,DimFei2003,agterberg2007magnetic} as well as to various superconducting magnetoelectric effects, such as inducing supercurrents with a static magnetization and vice-versa \cite{edelstein1995magnetoelectric, yip2002two,edelstein2005magnetoelectric,dimitrova2007theory,agterberg2012magnetoelectric,baumard2020interplay}. These effects are completely analogous to SGE/IEE and ISGE/EE in the normal state, respectively \cite{konschelle2015theory}. These phenomena are a basis of the emerging field of superconducting spintronics \cite{linder2015superconducting, eschrig2011spin}.
 
Another manifestation of SOC in the normal state is the weak antilocalization \cite{hikami1980spin, knap1996weak, wenk2010dimensional}. Namely, in metals with weak SOC, constructive electron interference along time-reversed trajectories increases the probability of electrons moving in closed loops. As a consequence, the conductance will be smaller compared to the classical (Drude) one. This phenomenon in known as weak localization. In the presence of strong SOC,  precession of electrons' spin leads to a phase shift, and consequently destructive interference and an increase in the Drude conductance. This is known as weak antilocalization, and it is a widely used tool for experimental characterization of SOC  \cite{koga2002rashba, bergmann1984weak, miller2003gate}.

 More recently, an equivalence between the singlet-triplet dynamics in diffusive superconductors and the spin-charge  transport in the normal state has been established ~\cite{bergeret2015theory,konschelle2015theory,tokatly2017usadel}.    In the linearized regime both phenomena are described by the same diffusion equation\cite{burkov2004theory, shen2014theory, raimondi2012spin}, the linearized Usadel equation.  The SOC enters this equation as spin  precession/relaxation terms,  and as charge-spin coupling term\cite{mal2005spin,stanescu2007spin,duckheim2009dynamic}, which in the superconducting case translate into a triplet-component precession and the  singlet-triplet coupling\cite{konschelle2015theory}. 
 Furthermore, weak localization is described in terms of two-particle correlation functions called Cooperons, which can also be obtained from these equations \cite{rammer2011quantum, knap1996weak, wenk2010dimensional} (see also Sec.~\ref{sec5}). Therefore, the linearized Usadel equation provides a universal quasiclassical description of the magnetoelectric phenomena in both normal and superconducting state, as well as weak localization, in the diffusive limit. In the opposite,  pure ballistic,  limit,  the system is described  by the  Eilenberger equation\cite{eilenberger1968transformation}. Its utility to study the triplet precession mediated by SOC  in ballistic superconducting systems has already been demonstrated in Refs.~\onlinecite{bergeret2014spin,konschelle2016ballistic,konschelle2016semiclassical}, whereas the singlet-triplet coupling has been analyzed in Ref.\onlinecite{konschelle2015theory} in the linearized case. In this work, we generalize all these works by providing the universal description of said phenomena at any disorder from the linearized Eilenberger equation.

 We  focus  on both the normal and superconducting state with arbitrary degree of disorder and discuss, based on the Eilenberger equation,  several applications related to spin transport and weak localization. As we will see, this equation provides a simple and physically transparent picture and allows for analytical solutions in many cases, while at the same time allowing to describe the full crossover from the diffusive to the ballistic limit. Moreover, we discuss the  one-to-one analogy to the singlet-triplet dynamics in the superconducting state as well as the appearance of non-conventional pair correlations induced by the SOC, which emerges naturally from the linearized Eilenberger equation. Our method can be easily adapted to different experimental setups, both in normal and superconducting regimes, as well to arbitrary linear in momentum SOC.

The article is organized as follows. First, in Sec.~\ref{sec2}, we introduce the linearized Eilenberger equation for systems with any linear-in momentum spin-orbit coupling, which is the central equation of this work,  and discuss the solution procedure in a general case,   for  an arbitrary  source term.  In Sec.~\ref{sec3}, we obtain a closed-form solution for the particular case of Rashba SOC\cite{bychkov1984properties}. We use this solution for three applications: local spin injection (Sec.~\ref{sec4}), superconducting Edelstein effect at arbitrary disorder (Sec.~\ref{sec5a}), and weak localization beyond the diffusive limit (Sec.~\ref{sec5}). In Sec.~\ref{sec6}, we solve the Eilenberger equation for the case of pure gauge SOC, and discuss spatial spin structures that form upon local spin injection. 
\section{The linear Eilenberger Equation and its general solution} 
\label{sec2}
We consider a system of conducting electrons with arbitrary  linear-in-momentum SOC, $H_{SO}=\alpha_k^ap_k\sigma^a$, where $p_k$ are components of the electron momentum, $\sigma^a$ are Pauli matrices, and $\alpha_k^a$ is a pseudotensor parametrizing a coupling of orbital and spin degrees of freedom. The system can be conveniently described using the SU(2) covariant\cite{mineev1992electric, frohlich1993gauge, jin20062,tokatly2008equilibrium} Hamiltonian
\begin{equation}
H=\frac{(p_k-\mathcal{A}_k)^2}{2m}+V_{imp},
\label{eq1}
\end{equation}
where $\mathcal{A}_k=\frac{1}{2} \mathcal{A}_k^a\sigma^a\equiv -m\alpha_k^a\sigma^a$ is an effective SU(2) vector potential, and the $V_{imp}$ accounts for random spin-independent disorder. In the superconducting state, the Hamiltonian \eqref{eq1}  acquires a structure in the Nambu space and needs to be supplemented with the superconducting pairing term which is off-diagonal in this space.

Within the quasiclassical approximation, which assumes that  all energy scales are much smaller than the Fermi energy $E_F$, our system is described by the two-times quasiclassical Green's function $\check{g(\mathbf{n},\mathbf{r},t,t')}$ in Keldysh-Nambu-spin space. It depends on the momentum direction $\mathbf{n}=\mathbf{p}/p_F$ and position $\mathbf{r}$, and satisfies the  Eilenberger equation
\begin{multline}
 v_F  n_i\tilde{\nabla}_i \check{g}+[\check{\omega}-i\check{\Delta},\check{g}]=\frac{1}{2m}\{n_i \mathcal{F}_{ij},\partial_{n_j}\check{g}\}+\frac{1}{2\tau}[\check{g},\langle \check{g} \rangle].
 \label{eq1a}
\end{multline}
In the absence of SOC, Eq.~\eqref{eq1a} can be derived by following the standard procedure\cite{eilenberger1968transformation,belzig1999quasiclassical}. However, for a correct inclusion of the SOC within the quasiclassical approach it is necessary to use the SU(2) covariant formulation, in which the SOC enters as a background SU(2) gauge field
~\cite{gorini2010non,bergeret2014spin, bergeret2015theory,tokatly2017usadel}. {Within this formulation  the Eilenberger equation is written in terms of covariant derivatives $\tilde{\nabla}_i \cdot=\partial_i-i[\mathcal{A}_i,\cdot]$ and the SU(2) magnetic field}   $\mathcal{F}_{ij}=\partial_i \mathcal{A}_j-\partial_j \mathcal{A}_i-i[\mathcal{A}_i,\mathcal{A}_j]$. ${1/\tau}$ is the disorder scattering rate, and $\langle...\rangle $ is the average over the direction of the Fermi momenta described by the  unit vector $\mathbf{n}$. Summation over repeated indices is implied.  
The commutator in the covariant derivative describes the spin-precession due to SOC, while the anticommutator in Eq.\eqref{eq1a} leads to singlet-triplet/spin-charge coupling.
Superconducting order is described by the anomalous self-energy term $\check{\Delta}$ and $\check{\omega}=\partial_t \delta(t-t')\tau_z$, where $\tau_i$ are Pauli matrices spanning the Nambu space. The Green's function has the following structure in the Keldysh subspace $\check{g}=\begin{bmatrix} g^{R} & g^{K} \\ 0 & g^A \end{bmatrix}$, where $R,A,K$ denote the retarded, advanced and Keldysh components, respectively.

 We first focus on  the normal state, $\Delta=0$, in which  $\check{g}$ is diagonal in the Nambu space, and 
the advanced and retarded components are trivial, $g^{R,A}(t,t')=\pm \delta(t-t')\tau_z$.   The  properties of the system are then solely determined by the non-equilibrium distribution function $f$ which is a matrix in spin space equal to the  Keldysh component evaluated at same times, $f(t)=\tau_z g^K(t,t)/2$. 
Then, starting from Eq.~\eqref{eq1a}, after performing the Fourier transform in the time domain, the Eilenberger equation reduces to 
\begin{multline}
 v_F  n_i\tilde{\nabla}_i f+i \omega f=\frac{1}{2m}\{n_i \mathcal{F}_{ij},\partial_{n_j}f\}\\-\frac{1}{\tau}(f-\langle f \rangle)+G(\mathbf{r}),
 \label{eq2}
\end{multline}
where $\omega$ is the frequency. In the right-hand side we have added a generic source term, $G({\rm r})$. Physically the latter describes a generation/injection of spin and/or charge. One possible realization of such term is a spin injection induced by a time-dependent Zeeman field $h$, as discussed in Ref. \cite{tokatly2017usadel}, for which $G({\rm r})\propto \partial h({\rm r})/\partial t \mapsto i\omega h$.
The distribution function $f$ in Eq. (\ref{eq2}) has the form
\begin{equation}
    f=f_0+f_j\sigma_j\; , \label{eq:structure_f}
\end{equation}
where $f_0$ describes  the non-equilibrium charge, and $f_j$, with $j=x,y,z$, the three non-equilibrium spin components. The anticommutator in the first term on the right-hand side of Eq.~\eqref{eq2} is responsible for the charge-spin coupling via the SU(2) magnetic field, which was widely studied in the context of the spin Hall effect \cite{gorini2010non,raimondi2012spin}.

One interesting aspect of  Eq. (\ref{eq2}) is that it also describes, after minor modifications, the  equilibrium properties  of either  a superconductor  at a temperature close to its critical temperature, or of  a  non-superconducting material weakly coupled to a superconductor.  
Being in equilibrium, these two situations can be written in terms of the Matsubara  frequencies $\omega_n=2\pi T (n+\frac{1}{2})$, {where $T$ is the temperature,} such that $\check g$ in  Eq.~\eqref{eq1a} is the quasiclassical Matsubara Green`s function which is a matrix  in the Nambu-spin space.  Eq. (\ref{eq1a}) has the same form after  substituting $\check{\omega}$ by $\omega_n$. 
Because  superconducting correlations are assumed to be weak, one  can  approximate $\check{g} \approx \begin{bmatrix} \text{sgn} \omega_n & f \\ \bar f & -\text{sgn} \omega_n \end{bmatrix} $, where now $f$  describes the superconducting  anomalous component of the Green's function and $\bar f$ its time-reversal conjugate, defined as $\bar f({\bf n})=\sigma_yf
^*(-{\bf n})\sigma^y$.  Linearization of the  the Eilenberger equation \eqref{eq1a} with respect to $f$  leads to Eq.~\eqref{eq2}  with the substitution $i \omega\to 2\omega_n$ and $(1/\tau)\to (1/\tau) \text{sgn} \omega_n$.

 The spin structure of $f$ is the same as in the normal case, Eq. (\ref{eq:structure_f}), but now $f_0$ describes the singlet component of the superconducting condensate, whereas $f_j$ describe the three triplet components.   Hence, the term with the SU(2) magnetic field in Eq. \ref{eq2} describes the singlet-triplet coupling via the SOC.  This establishes the equivalence between the spin-charge dynamics in the normal state with the singlet-triplet dynamics in the superconducting state - both are described by the linearized Eilenberger equation. This equivalence turns out to be very useful in tackling  transport  problems  of rather different systems and finding analogies between them, as we discuss in subsequent sections. But first, we present the general solution of the linear Eilenberger equation,  Eq.~\eqref{eq2}, which can be be applied to a wide range of problems.

In order to solve Eq.~\eqref{eq2}, we transform it to the momentum space, where $\mathbf{Q}$ is the momentum conjugated to the position $\mathbf{r}$, so we have 
\begin{multline}
f (1+iQ_i n_i l+i\omega\tau)-i[l n_i \mathcal{A}_i,f]\\=\langle f\rangle+G(\mathbf{Q})\tau+\frac{l}{2p_F}\{n_i \mathcal{F}_{ij}, \partial_{n_j}f\}\; , 
\label{eq3}
\end{multline}
where  $l=v_F \tau$ is the mean free path.
The second term in the first line  describes spin precession due to the SOC, whereas the last term  is the spin-charge coupling term.  The latter is a factor ${\cal A}/p_F$ smaller than the precession one. Therefore, within the quasiclassical approximation,  it can be treated perturbatively by  expanding  $f\approx f^{(0)}+f^{(p)}$, where the indices $0$ and $p$ denote the bare solution and the perturbative correction, respectively. Then, the following equations are satisfied 
\begin{equation}
f^{(0,p)}(1+i\delta+i\omega\tau)-[\Omega,f^{(0,p)}]=X^{(0,p)},
\label{eq4}
\end{equation}
where we  introduce the notation $\delta=Q_i n_i l$, $\Omega=l n_i \mathcal{A}_i$ and source terms  $X^{(0)}=\langle f \rangle+G(\mathbf{Q})\tau$ and $X^{(p)}=\frac{l}{2p_F}\{n_i \mathcal{F}_{ij},\partial_{n_j}f^{(0)}\}$. The solution of Eqs.~\eqref{eq4} can be written in terms of the averaged $\langle f\rangle$
\begin{multline}
f^{(0,p)}=\frac{1}{2|\Omega|^2}\frac{1}{1+i\delta+i\omega\tau} \{\Omega,X^{(0,p)} \}\Omega\\+\frac{i}{M}[\Omega, X^{(0,p)}]+\frac{1}{4|\Omega^2|}\frac{1+i\delta+i\omega\tau}{M}[\Omega,[\Omega,X^{(0,p)}],
\label{eq5}
\end{multline}
where $M=(1+i\delta+i\omega\tau)^2+4 |\Omega|^2$.

Finally,  we average Eq. (\ref{eq3}) over ${\bf n}$:
\begin{equation}
\bigg\langle (\delta+\omega\tau)f-[\Omega,f] \bigg\rangle=-iG(\mathbf{Q})\tau-i\langle X^{(p)}\rangle\; .
\label{eq6}
\end{equation}
This equation  determines $\langle f\rangle $ for any linear-in-momentum SOC. Once $\langle f \rangle$ is known, the full solution $f$ is readily obtained from Eqs.~\eqref{eq5}.   The  $\langle...\rangle$ average in Eq.~\eqref{eq6} can be performed analytically  in certain particular high-symmetry cases of SOC. In the present work  we will address two widely  studied cases:  Rashba SOC \cite{bychkov1984properties} in \ref{sec3} and the pure gauge SOC\cite{bernevig2006exact} in \ref{sec6}. In an arbitrary situation  Eqs.~\eqref{eq5}-\eqref{eq6} can be solved numerically. 

\section{Case of Rashba spin-orbit coupling \label{sec3}}

In this section, we provide the solution of the Eilenberger equation for the case of Rashba SOC, and discuss it in the diffusive limit (\ref{subsec3A}) and in the ballistic limit (\ref{subsec3B}).
The SU(2) vector potential for Rashba  SOC is given as $\mathcal{A}_x=-m \alpha \sigma_y$, $\mathcal{A}_y=m \alpha \sigma_x$, so that $|\Omega|=\alpha p_F \tau$. The SU(2) magnetic field $\mathcal{F}_{xy}$ is then $\mathcal{F}_{xy}=-\mathcal{F}_{yx}=-i[\mathcal{A}_x,\mathcal{A}_y]=2 m^2\alpha^2\sigma_z.$ To proceed, we expand $\langle {f}\rangle=\langle f_i \rangle\sigma_i$, $i=0,x,y,z$ . Then, starting from Eq.~\eqref{eq6}, after evaluating the averages over the Fermi surface, we can write a compact expression determining $\langle f\rangle$: \begin{equation}
\langle f_i \rangle =
[\hat{1}-{\Pi}(\mathbf{Q})]^{-1}_{ij}\Pi_{jk}(\mathbf{Q})G_k(\mathbf{Q})\tau.
\label{eq7}
\end{equation}
Here, $\hat{1}$ is the unity matrix, and $\Pi$ is the so-called matrix polarization operator defined  as
\begin{equation}
\Pi(\mathbf{Q})=
\begin{bmatrix}
a+b \cos 2\phi & b \sin 2 \phi & i e \cos \phi & i g \sin \phi \\
b \sin 2 \phi & a-b \cos 2\phi & i e \sin \phi & -i g \cos \phi \\
-i e \cos \phi & -i e \sin \phi & c & 0 \\
i g \sin \phi & -i g \cos \phi & 0 & d\\
\end{bmatrix},
\label{eq8}
\end{equation}
where $\phi$ is the angle associated with the momentum direction, such that $\cos\phi=Q_x/Q$. The first, second, third and fourth row/column in this matrix correspond to indices $x,y,z$ and $0$, respectively. The coefficients $a,b,c,d,e$ can be expressed as
\begin{align}
&a=\frac{1}{2T_0}+\sum_{\pm}\frac{1}{4T_\pm}, \qquad c=\sum_{\pm}\frac{1}{2T_{\pm}}, \qquad d=\frac{1}{T_0},\nonumber \\
&b=\frac{1}{2Q^2l^2}\bigg[\frac{(1+i\omega\tau)^2}{T_0}+T_0-\sum_{\pm}\frac{1}{2}\bigg(\frac{t_{\pm}^2}{T_{\pm}}+T_{\pm}\bigg)\bigg], \nonumber \\
&e= \sum_\pm \frac{\pm i t_\pm}{2Ql T_\pm}, \quad g=\frac{\gamma}{2Ql}\bigg[\frac{2x_\alpha}{T_0}(1+i\omega\tau)+i\sum_{\pm}\pm T_\pm\bigg].
\label{eq9}
\end{align}
Here, we used the notation $t_{\pm}=1\pm  i x_\alpha+i\omega \tau$, $T_{\pm}=\sqrt{t_{\pm}^2+Q^2l^2}$, $T_0=\sqrt{(1+i\omega\tau)^2+Q^2l^2}$, where $x_\alpha=\tau_\alpha^{-1}\tau$, with $\tau_\alpha^{-1}=2 p_F \alpha$ being the spin-orbit precession rate. Moreover, we introduced the quantity $\gamma=1/(2 p_Fv_F\tau)$.
The quantities $a,b,c$ and $d$ describe diffusion and relaxation processes, $e$ describes the inhomogeneous spin precession, while $g$ accounts for spin-charge/singlet-triplet coupling. Note that, when using Eqs.~\eqref{eq7},\eqref{eq8} and \eqref{eq9} to describe the superconducting state, we need to make the substitutions $i\omega\to 2\omega_n$ and $(1/\tau)\to (1/\tau)\text{sgn}\omega_n$, as explained in Sec.~\ref{sec2}. The later substitution also implies $l\to l\, \text{sgn} \omega_n$, $x_\alpha\to x_{\alpha} \,\text{sgn}\omega_n$, and $\gamma\to \gamma \,  \text{sgn}\omega_n$. 

It is important to emphasize that Eq. \eqref{eq7} is a compact way of writing $\langle f \rangle$, but it should bear in mind that the spin-charge coupling, {\it i.e.} the components proportional to $g$ in Eq. \eqref{eq8}, are treated perturbatively, and hence $\langle f \rangle$ contains term up to linear order in $g$. 

The operator $[\hat{1}-\Pi({\bf Q})]$ is the generalized diffusion operator which includes the charge-spin coupling term.  At $\mathbf{Q}$=0 and $\omega=0$ describes the relaxation properties of the system for arbitrary disorder:
\begin{equation}
\hat{1}-\Pi(\mathbf{Q}=0)=\text{diag}\bigg(\frac{1}{2}\frac{x_\alpha^2}{1+x_\alpha^2},\frac{1}{2}\frac{x_\alpha^2}{1+x_\alpha^2},\frac{x_\alpha^2}{1+x_\alpha^2},0\bigg),
\label{eq10}
\end{equation}  
where "diag" denotes a diagonal matrix. As expected, the charge component $\langle f_0\rangle$ is the only one which does not relax, in accordance to the charge conservation. In the diffusive limit, $x_\alpha\ll 1$, the relaxation operator yields the well known Dyakonov-Perel \cite{dyakonov1972spin} rates for Rashba SOC \cite{bychkov1984properties,manchon2015new,vzutic2004spintronics} (see Sec.~\ref{subsec3A}).

\subsection{Diffusive limit \label{subsec3A}}
In the  diffusive limit, the mean free path $l$ is the shortest length scale in the system. Therefore,  we can assume $Ql\ll 1$ and expand the quantities in Eq.~\eqref{eq9} keeping terms up to second order in $Ql$, and taking that SOC is weak compared to disorder ($x_\alpha\ll 1$). Eq.~\eqref{eq7} in this limit reduces to
\begin{equation}
\langle f_i\rangle=[\hat{1}-\Pi_D(\mathbf{Q})]^{-1}_{ij}G_j(\mathbf{Q})\tau,
\label{eq13a}
\end{equation}
where the diffusive polarization operator $\Pi_D(\mathbf{Q})$ is given by
\begin{multline}
\hat{1}-\Pi_D(\mathbf{Q})=\tau\hat{1}(i\omega+DQ^2)+\\\tau
\begin{bmatrix}
\tau_{DP}^{-1} & 0 & i\Gamma_p Q_x & i \Gamma_{sc} Q_y \\
0 & \tau_{DP}^{-1} & i \Gamma_p Q_y & - i \Gamma_{sc} Q_x \\
-i \Gamma_p Q_x & -i \Gamma_p Q_y & 2\tau_{DP}^{-1} & 0 \\
i \Gamma_{sc} Q_y &  - i \Gamma_{sc} Q_x & 0 & 0\\
\end{bmatrix}.
\label{eq11}
\end{multline}
Eqs.~\eqref{eq13a} and \eqref{eq11} yield the well known spin-charge coupled system of diffusion equations \cite{burkov2004theory, shen2014theory, raimondi2012spin}. Here, $D=\frac{1}{2}v_F^2\tau$ is the diffusion constant, $\tau_{DP}^{-1}=x_\alpha^2/(2\tau)$ is the Dyakonov-Perel spin relaxation rate,   $\Gamma_p=x_\alpha/\tau$ is the spin precession rate, and  $\Gamma_{sc}=\gamma x_\alpha^{3}/(2\tau)$ is the spin-charge coupling rate.
\subsection{Ballistic limit \label{subsec3B}}
If the mean free path is the longest length scale in the system, we may take the limit $l/l_\alpha \to \infty$, where $l_\alpha=v_F\tau_\alpha$. Eq.~\eqref{eq7} then becomes
\begin{equation}
\langle f_i \rangle=
[\Pi_B(\mathbf{Q})]_{ij}G_j(\mathbf{Q})\tau_\alpha,
\label{eq12}
\end{equation}
where we defined the ballistic polarization operator $\Pi_B(\mathbf{Q})$ as
\begin{equation}
\Pi_B(\mathbf{Q})=\lim_{l/l_\alpha->\infty}\frac{l}{l_\alpha} \Pi(\mathbf{Q}).
\label{eq13}
\end{equation}
$\Pi_B(Q)$ has the same form as $\Pi(\mathbf{Q})$ in Eq.~\eqref{eq8}, with the substitution $a\to a_B$, $b\to b_B$, $c\to c_B$, $d\to d_B$, $e\to e_B$, and $g\to g_B$. These coefficients acquire a particularly simple form at $\omega=0$, when they are purely real and read 
\begin{align}
& c_B=\Re \frac{1}{\sqrt{Q^2l_\alpha^2-1}},\, d_B=\frac{1}{\sqrt{Q^2l_\alpha^2+l_\alpha^2/l^2}},   \nonumber \\
& a_B=\frac{d_B}{2}+\frac{c_B}{4}, b_B=\frac{d_B}{2}-\frac{c_B}{4}+\frac{c_B}{Q^2l_\alpha^2}-2\delta(Ql_\alpha), \nonumber \\
& 
 e_B=-\frac{c_B}{Q l_\alpha}, \, g_B=\frac{\gamma_\alpha}{Q l_\alpha} \Im \frac{1}{\sqrt{Q^2l_\alpha^2-1}}.
 \label{eq14}
\end{align}
Here, we introduced the quantity $\gamma_\alpha=1/(2p_F v_F \tau_\alpha)$. Note that we keep the $l_\alpha/l$ contribution in the expression for $d_B$ in Eq.~\eqref{eq14}. Namely, if we neglected this contribution, we would have $d_B(Q)=1/(|Q|l_\alpha)$, which does not have a well defined Fourier transform  to the real space. Keeping small $l_\alpha/l$ regularizes the Fourier integral, which scales as $d_B(x)\sim\ln l_\alpha/l$.
{This fact will be used in Sec.~\ref{sec4} to obtain the results in the ballistic limit presented in Figs.~\ref{graphx} and \ref{graphy}.}

\section{Application: Local spin injection\label{sec4}}
\begin{figure}[h!]
	\includegraphics[width=0.4\textwidth]{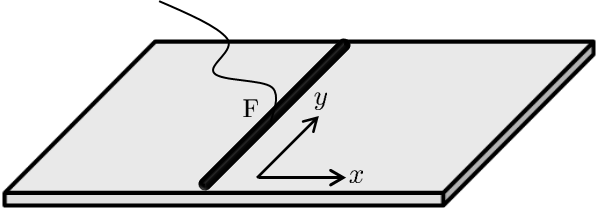}
	\caption{\label{schematic} Schematic representation of the spin injection experiment. The injector $F$ is a ferromagnetic electrode, oriented along the $y$-direction. }
\end{figure}

Having established the solution of the Eilenberger equation for the case of Rashba SOC  with an arbitrary source term in Sec.~\ref{sec3}, we now turn to various applications. In this section, we  consider the  problem of spin injection from  a  narrow, infinitely long ferromagnetic electrode, placed on top of a Rashba conductor (see Fig.~\ref{schematic}). The electrode lies along the $y$-direction, so the system is inhomogeneous only along the $x$-direction, which makes this problem effectively one-dimensional. Such spin injection can be modeled by a source term $G(\mathbf{r})=G_i\sigma_i\delta(x)$, where $i=x,y,z$ denotes the injected spin component. Conveniently, in the momentum space this source term reduces to a constant, $G_i(\mathbf{Q})=G_i$.

This kind of setup was already studied in Ref.~\onlinecite{burkov2004theory}, but only in the diffusive limit. By contrast, our result provides a full crossover from the ballistic to the diffusive limit. Furthermore, our approach based on the Eilenberger equation is significantly simpler compared to the  standard density matrix calculation employed in Ref.~\cite{burkov2004theory}.

The spin density $S_i$ ($i=x,y,z$) is given as
\begin{equation}
S_i=N_F\langle f_i \rangle,
\label{eq15}
\end{equation} 
where $N_F$ is the density of states at the Fermi energy. In the linearized Eilenberger equation, we do not treat explicitly the mean-field electrostatic potential by absorbing it into the definition of the charge distribution function $f_0$. In this approach the average value $\langle f_0\rangle$ yields the variation of the electrochemical potential
\begin{equation}
\delta\mu=\langle f_0\rangle.
\label{eq15-1}
\end{equation}
If needed, the corresponding charge density $\delta n = N_F(\delta\mu + e\delta\varphi)$ can be determined by solving the Poisson equation for the electrostatic potential $\delta\varphi$. 

The polarization operator in the present 1D setup ($\phi=0$) acquires the form  
\begin{equation}
\Pi(Q)=
\begin{bmatrix}
a+b & 0 & i e & 0 \\
0 & a-b & 0 & -i g  \\
-i e  & 0 & c & 0 \\
0 & -i g  & 0 & d\\
\end{bmatrix}.
\label{eq16}
\end{equation}

In the following, we will be interested in the spatial evolution of $S_i$ and $\delta\mu$, which is obtained by solving Eq.~\eqref{eq7} and performing the Fourier transform
\begin{equation}
f(x)=\mathcal{F}[f(Q)]=\int_{-\infty}^{\infty}\frac{dQ}{2\pi}e^{iQx}f(Q).
\label{eq17}
\end{equation}
The polarization of the injected spin can be controlled by changing the magnetization direction of the ferromagnetic electrode.  In the following, we will consider two scenarios: first, injection of the spin component perpendicular to the injector ($G_x$), and second, injection of the spin component parallel to the injector ($G_y$). They are addressed in Subsecs.~\ref{subsec4A} and \ref{subsec4B}, respectively.  The dynamics of spin and charge is governed by different mechanisms in the two scenarios. Namely, in the first scenario, the injected spin $S_x$ induces $S_z$ via the inhomogeneous spin precession. Coupling between these two spin components is described by the coefficient $e$ in the polarization operator $\Pi(Q)$ [see Eq.~\eqref{eq16}]. In the second scenario, the injected spin $S_y$ induces a charge density $\delta n$ via the spin-charge coupling [coefficient $g$ in $\Pi(Q)$]. 
\subsection{Injection of spin polarized in  $x$-direction \label{subsec4A}}
If the $x$-component of the spin is injected ($G_x\neq 0, G_y=G_z=0$), this leads to the finite $S_x$ component in the system, but also a finite $S_z$ component, induced by the inhomogeneous spin precession. Solving Eq.~\eqref{eq7} yields
\begin{align}
&\frac{S_x(Q)}{N_FG_x\tau_\alpha}=x_\alpha\frac{(a+b)(1-c)+e^2}{(-1+a+b)(-1+c)-e^2}, \nonumber\\
&\frac{S_z(Q)}{N_FG_x\tau_\alpha}=x_\alpha\frac{-i e}{(-1+a+b)(-1+c)-e^2}.
 \label{eq18}  
\end{align}
Furthermore, there is a finite charge current flowing in the $y$-direction, defined as
\begin{equation}
J_y(Q)=N_F\langle n_y f_0\rangle .
\label{eq19}
\end{equation}
Using the expression for $f_0$ obtained using Eq.~\eqref{eq5}, we find
\begin{align} 
&\frac{J_y(Q)}{\gamma_\alpha}=\frac{iS_z(Q)}{2Ql}\bigg[-2T_0+\sum_{\pm} \frac{t_\pm+Q^2l^2}{T_\pm}\bigg]\nonumber \\&-\frac{i(S_x(Q)+N_FG_x\tau)}{2 Q^2 l^2} \bigg[ix_\alpha T_0-\sum_{\pm}\bigg(\frac{i x_\alpha Q^2l^2}{T_\pm}\pm T_{\pm}\bigg)\bigg].
\label{eq20}
\end{align}
\paragraph{Total Spin and current} 
The total (integrated) spin for the component $S_x$ is readily found as 
\begin{equation}
\int_{-\infty}^{\infty} dx S_x (x)= S_x(Q=0)=N_Fx_\alpha\bigg(1+\frac{2}{x_\alpha^2}\bigg) G_x\tau_\alpha.
\label{eq21}
\end{equation}
Similarly, 
\begin{equation}
\int_{-\infty}^{\infty} dx S_z(x)=0, \quad \int_{-\infty}^{\infty} J_y(x)dx=N_Fx_\alpha G_x\tau_\alpha\gamma_\alpha.
\label{eq22}
\end{equation}
Note that $S_x$ and $J_y$ are even functions in $x$, whereas $S_z$ is an odd function. For that reason, integrated $S_z$ yields zero. 
\paragraph{Diffusive limit}
In the diffusive limit $x_\alpha\ll 1$, it is possible to obtain analytical expressions for $S_x(x)$, $S_z(x)$ and $J_y(x)$. Starting from the polarization operator specified in Eq.~\eqref{eq11}, after the Fourier transform we obtain 
\begin{align}
&\frac{ S_x(x)}{N_FG_x\tau_\alpha}=\Re\frac{1+\frac{5i}{\sqrt{7}}}{x_\alpha \kappa} e^{-\frac{\kappa|x|}{l_\alpha}}, \frac{ S_z(x)}{N_FG_x\tau_\alpha}=\frac{4x \Im e^{-\frac{\kappa |x|}{l_\alpha}}}{\sqrt{7}|x|x_\alpha}, \nonumber \\  
&\frac{J_y(x)}{N_FG_x\tau_\alpha \gamma_\alpha}=\Re  x_\alpha\frac{\frac{3i}{\sqrt{7}}-1}{\kappa}e^{-\frac{x}{l_\alpha}\kappa}.
\label{eq23}
\end{align}
where $\kappa^2=(-1+i\sqrt{7})/2$.
\paragraph{Ballistic limit}
Starting from Eq.~\eqref{eq12}, we straightforwardly obtain the following expressions in the ballistic limit
\begin{align}
&\frac{S_x(Q)}{N_FG_x \tau_\alpha}=a_B(Q)+b_B(Q),\qquad \frac{S_z(Q)}{N_FG_x \tau_\alpha}=-i e_B(Q), \nonumber \\ &\frac{J_y(Q)}{N_FG_x\tau_\alpha \gamma_\alpha}=a_B(Q)-2c_B(Q).
\label{eq24}
\end{align}

\begin{figure}[h!]
	\includegraphics[width=0.35 \textwidth]{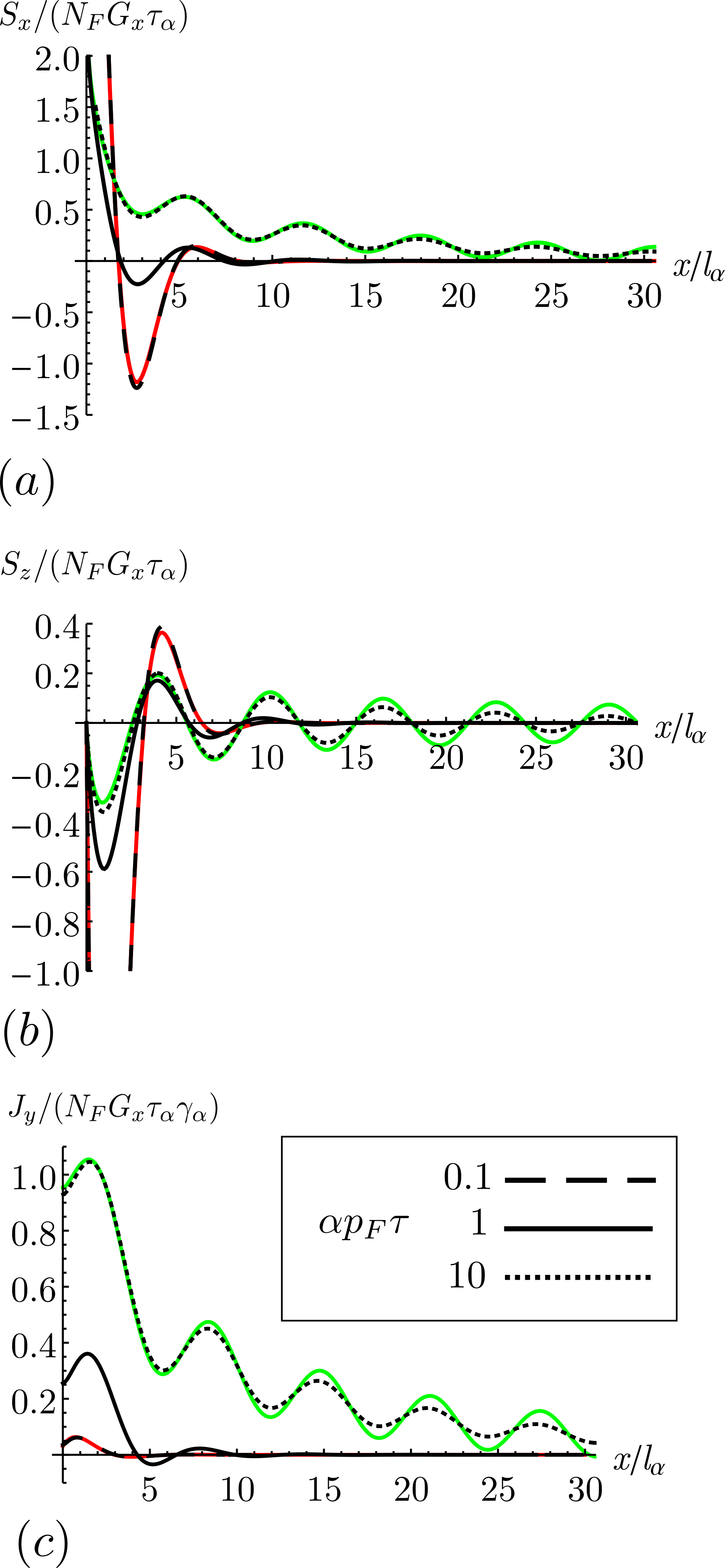}
	\caption{\label{graphx} Spatial dependence of spin density $S_x$  $(a)$,  spin density $S_z$  $(b)$, and charge current density $J_y$  $(c)$, induced by a local spin injection of the $x$-component of the spin. Black curves correspond to the exact numerical solution, whereas the red and green curves are the approximate solutions in the diffusive limit (for $\alpha p_F\tau=0.1$) and in the ballistic limit, respectively.}
\end{figure}

We calculate $S_x(x)$, $S_z(x)$ and $J_y(x)$ by performing the Fourier transform of Eqs.~\eqref{eq18} and \eqref{eq20} numerically. The results are shown in Fig.~\ref{graphx} as black curves for various values of $\alpha p_F\tau$. The expressions obtained in the diffusive and ballistic limit are also plotted as colored curves for comparison. All three quantities oscillate in space with a period determined by the spin precession length $l_\alpha$ , and decay on the distances comparable to the mean free path $l$ due to spin relaxation. 

\subsection{Injection of spin polarized in $y$-direction \label{subsec4B}}
Next, we consider injection of the $y$-component of the spin $(G_y\neq 0, G_x=G_z=0)$. Aside from the finite spin density $S_y$, a finite change of the electrochemical potential $\delta\mu$, leading to a finite charge density $\delta n$, is also generated, due to spin-charge coupling. Solving Eq.~\eqref{eq7}, we obtain
\begin{align}
&\frac{S_y(Q)}{N_F G_y\tau_\alpha}=x_\alpha\frac{b-a}{(-1+a-b)},\nonumber  \\
&\frac{\delta\mu(Q)}{G_y\tau_\alpha}=x_\alpha\frac{ig}{(-1+a-b)(-1+d)}.
\label{eq25}
\end{align}
Unlike previously considered case in Sec.~\ref{subsec3A}, there is no finite charge current - we readily check that $J_x(Q)=N_F\langle n_x f_0 \rangle=0$.

The electrochemical potential $\delta\mu(Q)$ has a pole of order 1 at $Q=0$. We may add and subtract $\text{Res}[\delta\mu(Q=0)]/(Ql)$ from $\delta\mu(Q)$, and apply the Fourier transformation \eqref{eq17}. This way, we obtain
\begin{equation}
\frac{\delta\mu(x)}{G_y \tau_\alpha}= \mathcal{F}\bigg[\delta\mu(Q) + \frac{2i\gamma_\alpha}{Q l} \bigg]+2\gamma_\alpha \Theta(x).
\label{eq26}
\end{equation}
This equation describes a voltage jump, from zero to the maximal value determined by the prefactor of the $\Theta$-function
\begin{equation}
{\delta\mu^{max}}=2G_y\tau_\alpha \gamma_\alpha = \frac{G_y}{p_Fv_F}.
\label{eq27}
\end{equation}
Therefore, the system acts as a spin-controlled battery: by injecting a $y$-component of a spin, a voltage drop is generated as a consequence of spin-charge conversion. Remarkably, the generated voltage drop of Eq.~\eqref{eq27} is universal and depends neither on SOC strength $\alpha$ nor on the momentum relaxation time $\tau$. Of course, this holds true only if the size of the sample in the $x$-direction is larger that the spin precession length and the mean free path.

\paragraph{Total Spin}
Similarly to Eq.~\eqref{eq21}, we find the total spin $S_y$ as
\begin{equation}
\int_{-\infty}^{\infty}dx S_y(x)=N_F x_\alpha \bigg(1+\frac{2}{x_\alpha^2}\bigg)G_y\tau_\alpha.
\label{eq28}
\end{equation}
\paragraph{Diffusive limit}
In the diffusive limit $x_\alpha\ll 1$, we readily obtain analytical results in the real space by utilizing the polarization operator in Eq.~\eqref{eq11}:
\begin{align}
&\frac{S_y(x)}{N_FG_y\tau_\alpha}=\frac{e^{-|x|/l_\alpha}}{x_\alpha},\nonumber \\ &\frac{ \delta\mu(x)}{G_y\tau_\alpha\gamma_\alpha}=e^{x/l_\alpha}\Theta(-x)-e^{-x/l_\alpha}\Theta(x)+2\Theta(x).
\label{eq29}
\end{align}
\paragraph{Ballistic limit}
Using  Eq.~\eqref{eq12}, we find in the ballistic limit
\begin{equation}
\frac{S_y(Q)}{N_FG_y\tau_\alpha}=a_B(Q)-b_B(Q), \qquad \frac{\delta\mu(Q)}{G_y\tau_\alpha}=-i e_B(Q).
\label{eq30}
\end{equation}

\begin{figure}[h!]
	\includegraphics[width=0.35\textwidth]{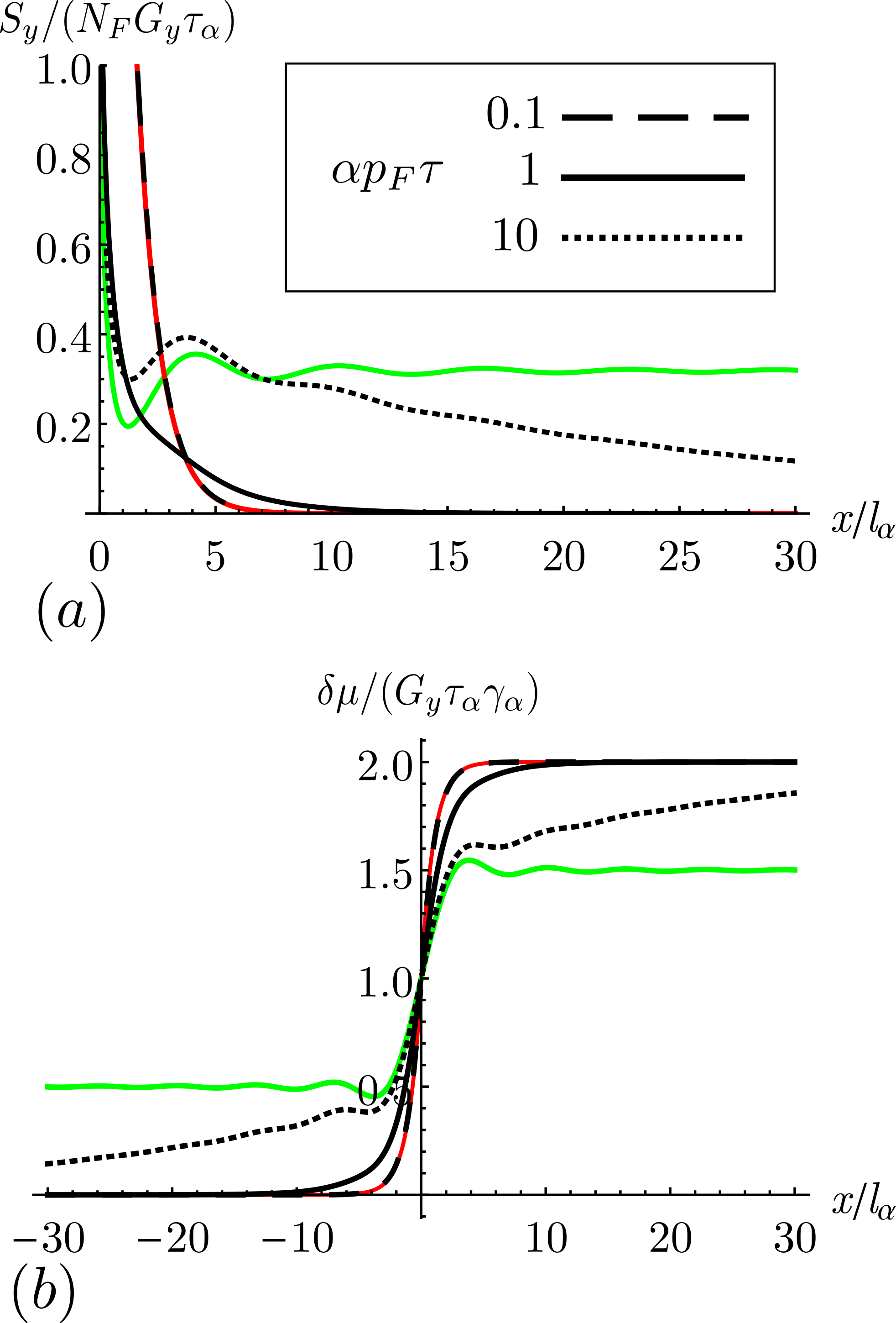}
	\caption{\label{graphy} Spatial dependence of spin density $S_y$  $(a)$, and electrochemical potential $\delta\mu$ $(b)$, induced by a local spin injection of the $y$-component of the spin. Black curves correspond to the exact numerical solution, whereas the red and green curves are the approximate solutions  in the diffusive limit (for $\alpha p_F\tau=0.1$) and in the ballistic limit, respectively.}
\end{figure}

We calculate $S_y(x)$ and $\delta\mu(x)$ by performing the Fourier transform \eqref{eq17}. The results are shown in Fig.~\ref{graphy} as black curves for various values of $\alpha p_F\tau$. The expressions obtained in the diffusive and ballistic limit are also plotted as colored curves for comparison.

\section{Application: Superconducting Edelstein effect at arbitrary disorder}\label{sec5a}
In his seminal work, Edelstein showed that an equilibrium supercurrent in Rashba superconductors can generate a finite spin polarization, both in the ballistic \cite{edelstein1995magnetoelectric} and the diffusive limit \cite{edelstein2005magnetoelectric}. This is known as the superconducting Edelstein effect, and it is naturally understood as a consequence of SOC-mediated singlet-tripled coupling \cite{konschelle2015theory}. In this section, we will apply the results of Secs.~\ref{sec2} and \ref{sec3} to study this effect. As we will show, our approach allows us to reproduce aforementioned Edelstein's results and generalize them for arbitrary disorder in just a few lines of calculation.
  
In the superconducting state, the  superconducting correlations appear due to the source term $G(\mathbf{r})=-2i\Delta e^{i\phi}\text{sgn} (\omega_n)$, which describes singlet s-wave pairing.  The superconducting phase with the form $\phi=\mathbf{q}\cdot \mathbf{r}$ gives the supercurrent flowing through the system,  $\mathbf{j}\sim \mathbf{q}$. For simplicity, we assume that $\mathbf{q}$ lies along the $x$-direction, $\mathbf{q}=(q,0)$. In momentum space  the source reads  $G(\mathbf{Q})=-4i\pi \Delta \delta(\mathbf{Q}-\mathbf{q}) \text{sgn}(\omega_n)$, and one can  solve Eq.~\eqref{eq7} straightforwardly after the following substitutions in Eqs.~\eqref{eq7} and \eqref{eq8}: $i\omega\to 2\omega_n$ and $(1/\tau)\to (1/\tau) \text{sgn}(\omega_n)$ (see  Sec.~\ref{sec2}). 

From Eq.~\eqref{eq7} we obtain a finite averaged singlet component $\langle f_0 \rangle$, and a triplet component $\langle f_y\rangle$ induced by spin-charge/singlet-triplet coupling. They are given by  
\begin{align}
&\langle f_0\rangle(\mathbf{Q})=-\frac{4i\pi\Delta \tau d}{1-d}\delta(\mathbf{Q}-\mathbf{q}),\nonumber \\
&\langle f_y\rangle(\mathbf{Q})=\frac{4\pi\Delta\tau g}{(-1+a+b)(d-1)}\delta(\mathbf{Q-q}).
\label{eq33a}
\end{align}
We are interested in the Edelstein effect, {\it i.e.} the linear response to the supercurrent. Thus, we may expand $\langle f_0\rangle$ and $\langle f_y\rangle$ from Eq.~\eqref{eq33a}, retaining only the terms linear in $q$:
\begin{equation}
\langle f_0 \rangle=-\frac{i\Delta e^{iqx}}{|\omega_n|}, \qquad \langle f_y\rangle=\frac{\Delta e^{iqx} \gamma ql}{\omega_n} \frac{x_\alpha \tilde{x}_\alpha^2}{\tilde{x}_\alpha^2+4|\omega_n|\tau},
\end{equation}
where we introduced $\tilde{x}_\alpha^2=x_\alpha^2/[x_\alpha^2+(1+2|\omega_n|\tau)^2]$. Note that the average singlet component $\langle f_0 \rangle$ is even in frequency, while the average triplet component $\langle f_y\rangle$ is odd in frequency\cite{bergeret2005odd}.

Using Eq.~\eqref{eq5}, we can now find the complete solution for the anomalous Green's function $f$. For the singlet component we obtain
\begin{equation}
f_0=\langle f_0 \rangle\bigg[1-\frac{iq l n_x \text{sgn}(\omega_n) }{1+2|\omega_n|\tau}\bigg].
\label{eq35a}
\end{equation}
Here the first and second contributions have an $s$-wave and $p$-wave symmetry, respectively. The triplet components are
\begin{align}
& f_y=\langle f_y\rangle\bigg[1+ \frac{2|\omega_n|\tau}{1+2|\omega_n|\tau}(-n_x^2+n_y^2)\bigg], \nonumber \\
&f_x=\langle f_y\rangle \frac{4|\omega_n|\tau}{1+2|\omega_n|\tau} n_xn_y, \quad f_z=\langle f_y\rangle \frac{4\omega_n \tau}{x_\alpha}n_y.
\label{eq36a}
\end{align}
The first contribution of $f_y$ has an $s$-wave symmetry, whereas  the second contribution of $f_y$ and $f_x$ have a $d$-wave symmetry. All these triplet components, being even in momentum, are due to Pauli's exclusion principle odd in frequency\cite{berezinskii1974new,bergeret2005odd,bergeret2007odd,eschrig2007symmetries,linder2019odd}, as one can check explicitly from the above expressions. In contrast, $f_z$ has a $p$-wave symmetry (odd in momentum) and hence it is an  even function of the Matsubara frequency. 

Finally, having found  $f$, we now proceed to calculate observables. For superconductors in the linearized regime, the spin polarization can be calculated from the expression\cite{konschelle2015theory}
\begin{equation}
S_i=\frac{i\pi}{4}N_F T \sum_{\omega_n} \text{Tr}\langle \sigma_i (f\bar{f}+\bar{f}f) \rangle\text{sgn}(\omega_n).
\end{equation}
Substituting the solutions from Eqs.~\eqref{eq35a} and \eqref{eq36a}, keeping only the terms up to linear order in $q$, we obtain
\begin{equation}
S_y=\pi N_F T\sum_{\omega_n}\frac{\gamma q l \Delta^2}{|\omega_n|^2}\frac{x_\alpha \tilde{x}_\alpha^2}{\tilde{x}_\alpha^2+4|\omega_n|\tau}.
\end{equation} 
This is the main result of this section, describing the Edelstein effect in superconductors with arbitrary disorder. In the diffusive limit, it reduces the result of Ref.~\onlinecite{edelstein2005magnetoelectric}:
\begin{equation}
S_y=\pi N_F T \sum_{\omega_n}\frac{ql\Delta^2}{|\omega_n|}\frac{\Gamma_{sc}}{\tau_{DP}^{-1}+2|\omega_n|},
\end{equation}
whereas in the ballistic limit, we reproduce the result of Ref.~\onlinecite{edelstein1995magnetoelectric}
\begin{equation}
S_y=\pi N_F T\sum_{\omega_n}\frac{1}{4}\frac{q}{p_F}\frac{\Delta^2}{\omega_n^2}\frac{(\alpha p_F)^3}{|\omega_n|[(\alpha p_F)^2+\omega_n^2]}.
\end{equation}

In this section, we demonstrated that our solution of the linearized Eilenberger equation, presented in Sec.~\ref{sec2} and \ref{sec3}, can be a powerful tool in the study of magnetoelectric phenomena in superconductors at arbitrary disorder. The same procedure  could be applied to study magnetoelectric effects in systems with different kinds of linear-in-momentum SOC (other than Rashba).

\section{Application: Weak localization beyond the diffusive limit \label{sec5}}
Theory of WAL in a Rashba electron gas is well established in the diffusive limit $x_\alpha\ll 1$ \cite{knap1996weak,wenk2010dimensional, iordanskii1994weak}.  More recently, Refs.~\onlinecite{araki2014weak} and \onlinecite{guerci2016spin} attempted to extend this theory beyond the diffusive limit ($x_\alpha\sim1$). However, their results are not correct due to the inadequate $Q$ expansion of the two-particle correlators  (Cooperons) that determine the W(A)L, as we can easily check using our method  and  we discuss  in detail in the following. 

 WL corrections are most commonly studied using the diagrammatic perturbation theory, which involves calculating disorder averages of two Green's functions corresponding to maximally crossed diagrams called Cooperons \cite{akkermans2007mesoscopic}. In this work, we will use a different approach, which is more physically transparent and more easily employed beyond the diffusive limit. Namely, we will exploit the fact that the resolvent of the linearized Eilenberger equation also leads to the Cooperon \cite{rammer2011quantum}. This holds because superconducting (particle-hole) correlations of the linearized Eilenberger equation are equivalent to maximally crossed diagrams. This approach is similar to the field-theoretical treatment of WL using the non-linear $\sigma$ model \cite{kamenev2011field, hikami1981anderson}

In order to calculate the weak localization correction to the conductance in the 2D Rashba conductor, we start from the main building block - the Cooperon. It is given as
\begin{equation}
C^{-1}(\mathbf{Q})=2\pi N_F \tau[1-\Pi(\mathbf{Q}],
\label{eq31}
\end{equation}
 where $\Pi(\mathbf{Q})$ is the polarization operator defined in Eq.~\eqref{eq8}. Then, the interference correction to the Drude conductance is
\begin{equation}
\delta\sigma=\frac{e^2}{2\pi}\int \frac{d^2\mathbf{Q}}{(2\pi)^2}\text{Tr}[C(\mathbf{Q})W].
\label{eq32}
\end{equation}
Here, $W$ is the so-called Cooperon weight factors, given as $W=\text{diag}(W_x,W_y,W_z,W_0)$, where
\begin{align}
&W_0=2\pi N_F v_F^2\tau_0^3, \, W_x=-W_0\bigg(\frac{1+\frac{x_\alpha^2}{4}}{1+x_\alpha^2}-\frac{x_\alpha^2}{2(1+x_\alpha^2)^2}\bigg), \nonumber \\
&W_y=-W_0\frac{1+\frac{3x_\alpha^2}{4}}{1+x_\alpha^2}, \,W_z=-W_0 \bigg(\frac{1}{1+x_\alpha^2} -\frac{x_\alpha^2}{2(1+x_\alpha^2)^2}\bigg).
\label{eq33}
\end{align}
Eqs.~\eqref{eq32} and \eqref{eq33} are proved in Appendix \ref{app1} using the diagrammatic perturbation theory.

After inverting Eq.~\eqref{eq31} and integrating over the angle $\phi$, we obtain $C(Q)=\frac{1}{2\pi}\int_0^{2\pi}\text{d} \phi\, C(\mathbf{Q})=\text{diag}(C_x,C_y,C_z,C_0)$, where
\begin{multline}
C_{x,y}=\bigg[\frac{(4\pi N_F\tau)^{-1}}{1-(a-b)}+\frac{(4\pi N_F\tau)^{-1}(1-c)}{[1-(a+b)](1-c)-e^2}\bigg],\\
C_z=\frac{(2\pi N_F\tau)^{-1}[1-(a+b)]}{[1-(a+b)](1-c)-e^2}, \, C_0=\frac{(2\pi N_F\tau)^{-1}}{1-d}.
\label{eq34}
\end{multline}
The Cooperons $C_{x,y,z}$ correspond to the three spin-triplets, while $C_0$ is the spin singlet. The weight factor for the singlet channel $W_0$ is always positive, meaning that it contributes as a positive (antilocalization) contribution to $\delta \sigma$. In contrast, the triplet weight factors $W_{x,y,z}$ are always negative, and yield a negative (localization) correction to $\delta\sigma$. In the diffusive limit $x_\alpha\ll 1$, we recover the well known result $W_{x,y,z}=-W_0$ \cite{knap1996weak}.  

Note that in Eq.~\eqref{eq34} we have neglected the effect of spin-charge coupling, given by the coefficient $g$ in the polarization operator $\Pi$. This is because in the quasiclassical approximation $e\sim \alpha/v_F\ll 1$, so it gives a negligible contribution compared to other parameters. Therefore, the singlet Cooperon $C_0$ is unaffected by spin-orbit coupling. An interesting open question is whether for  $\alpha\sim v_F$ spin-charge coupling leads to a suppression of the singlet $C_0$.

To proceed, we note that the integral in Eq.~\eqref{eq32} is dominated by the small $Q l$ values. Therefore,  we may expand the coefficients $a,b,c,d,e$ assuming small $Q l$, keeping terms up to fourth order: $a(Q)\approx a_0+a_2 Q^2+a_4 Q^4, b(Q)\approx b_2 Q^2+b_4 Q^4, c(Q)\approx c_0+c_2 Q^2+c_4 Q^4, d(Q)\approx d_0+ d_2 Q^2+d_4 Q^4,$ and $e(Q)=e_ 1 Q+e_3 Q^3$. All expansion coefficients are defined in Appendix \ref{app1}. Then, in the denominators of the first contribution of $C_{x,y}$ and in $C_0$ we keep terms up to second order in $Q$. In denominators of the second contribution in $C_{x,y}$ and in $C_z$ we keep terms up to fourth order in $Q$, while we keep terms up to second order in the numerators. This way, all Cooperons can be expressed as diffusion poles after performing a partial fraction decomposition, namely
\begin{align}
&C_{x,y}=\sum_{i=1,2,3}\frac{(2\pi N_F\tau^3v_F^2)^{-1}A_{xi}}{Q^2+\lambda_i^{-2}}, \nonumber \\ &C_z=\sum_{i=1,2} \frac{(2\pi N_F\tau^3v_F^2)^{-1}A_{zi}}{Q^2+\lambda_i^{-2}}, \, C_0=\frac{(2\pi N_F\tau^3 v_F^2)^{-1}}{\frac{1}{2}Q^2}.
\label{eq35}
\end{align}  
Here, we introduced the relaxation lengths $\lambda_i$, which is specified in the Appendix~\ref{app2} together with the coefficients $A_{x,zi}$. 

In the study of weak localization it is customary to stop the $Ql$ expansion of the Cooperon at the second order, as was indeed done in Refs.~\cite{guerci2016spin} and \cite{araki2014weak}. However, we find that this is justified only in the diffusive limit $x_\alpha\ll 1$. Beyond this limit, it is actually important to keep the terms up to fourth order in $Ql$, as they are needed to obtain the correct value of relaxation lengths $\lambda_1$ and $\lambda_2$. This can be seen from the explicit equation for $\lambda_{1,2}$  in Appendix \ref{app2}. Here, for $x_\alpha\gtrsim 1$, we see that the higher-order expansion coefficients $a_4,b_4,c_4$ and $e_3$ give contributions of the same order of magnitude as the lower-order expansion coefficients $a_{0,2},b_{2},c_{0,2}$ and $e_1$. 

\begin{figure}[h!]
	\includegraphics[width=0.4\textwidth]{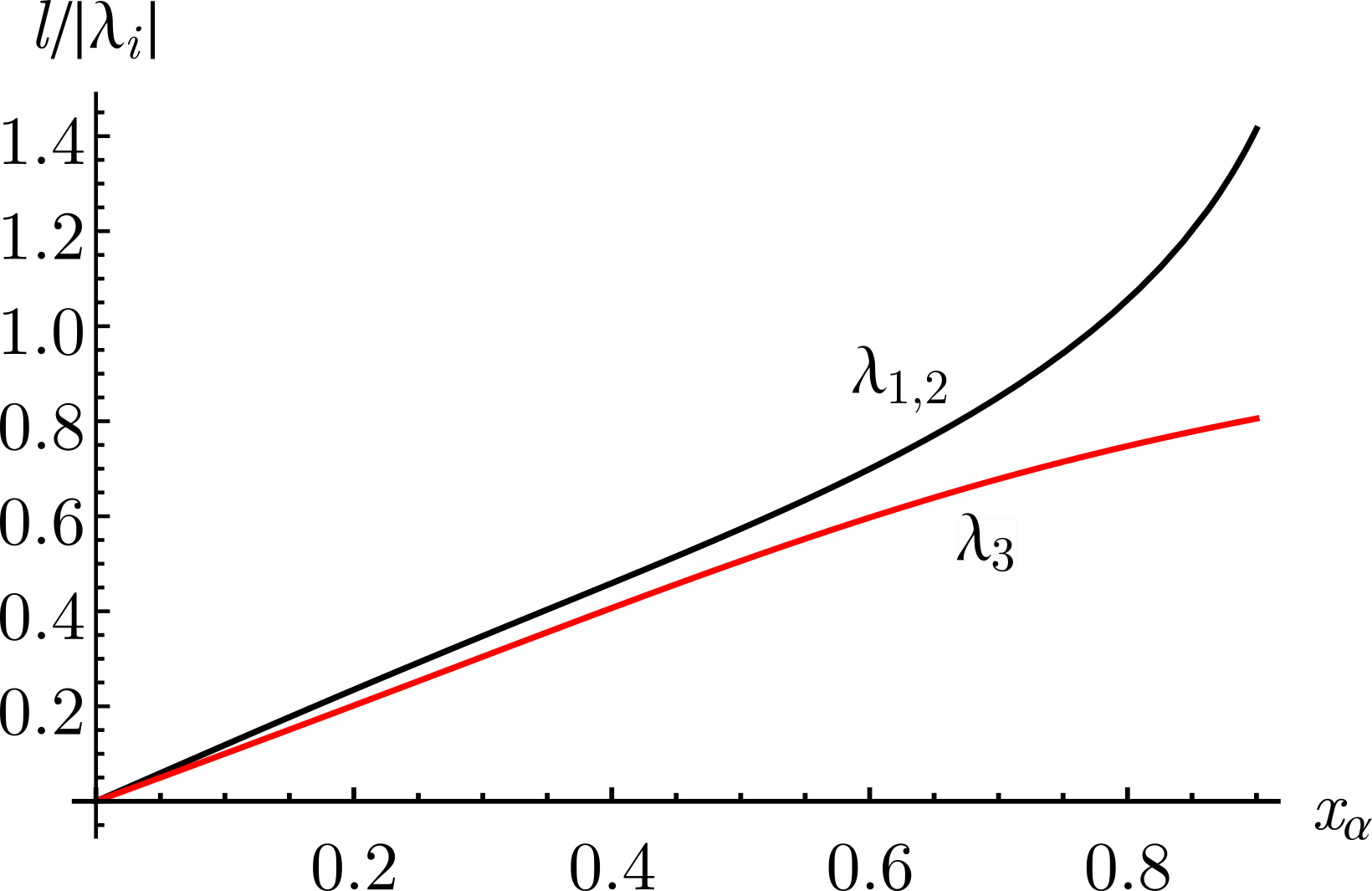}
	\caption{\label{figureWL1} Inverse relaxation lengths $|\lambda_i|^{-1}$ as a function of the parameter $x_\alpha$. The quantities $\lambda_1$ and $\lambda_2$ are complex conjugated in the plotted range, while $\lambda_3$ is real.}
\end{figure}

In Fig.~\ref{figureWL1} we plot the inverse relaxation lengths as a function of $x_\alpha$. The lengths $\lambda_{1,2}$ are complex, while $\lambda_3$ is real. In the strict diffusive approximation $x_\alpha\ll 1$, these lengths are given as $\lambda_{1,2}^{-1}=\sqrt{(-1\pm i\sqrt{7})/2l_\alpha^2}$ and $\lambda_3^{-1}=1/l_\alpha$ \cite{sanz2019nonlocal, burkov2004theory}.

\begin{figure}[h!]
	\includegraphics[width=0.4\textwidth]{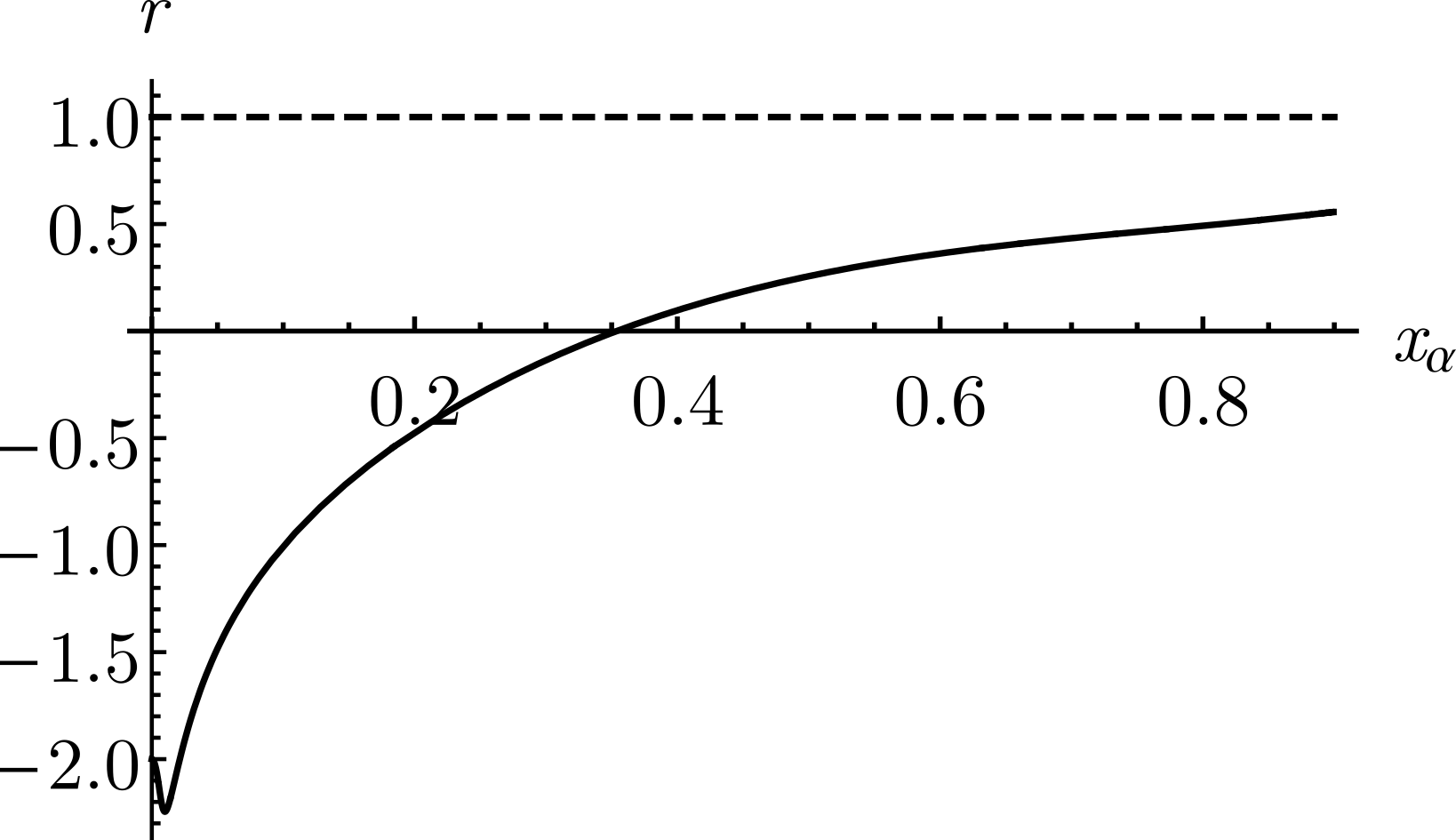}
	\caption{\label{figureWL2} Weak localization correction to the conductance normalized with respect to the singlet-channel contribution. The ratio $L/l$ is fixed to 100.}
\end{figure}

Finally, the WL correction to the conductance is
\begin{equation}
\delta\sigma=\frac{e^2}{2\pi}\int_{1/L}^{1/l}\frac{Q dQ}{2\pi}\sum_{i=0,x,y,z}W_iC_i(Q).
\label{eq36}
\end{equation}
Here, we introduce the upper and lower cutoff ot the integral in Eq.~\eqref{eq36}, determined by the inverse size of the system $(L)$ and the inverse mean free path ($l$), respectively. After performing the $Q$ integration, we have 
\begin{equation}
\frac{\delta\sigma}{\sigma_0}=\ln \frac{L}{l}+\sum_{i=1,2,3}K_i \ln \frac{1+\lambda_i^2/l^2}{1+\lambda_i^2/L^2},
\label{eq37}
\end{equation}
where we introduced $K_{1,2}=\frac{1}{4}[A_{x1,2}(W_x+W_y)+A_{z1,2}W_z]$, $K_3=\frac{1}{4}A_{x3}(W_x+W_y)$, and $\sigma_0=e^2/(2\pi^2)$ is the conductance quantum. The first term in Eq.~\eqref{eq37} comes from the singlet channel, which is not affected by the SOC, while all other terms come from triplet channels and are suppressed by the SOC. In Fig.~\ref{figureWL2}, we plot the WL conductance normalized with respect to the singlet contribution 
\begin{equation}
r=\frac{\delta\sigma}{\sigma_0 \ln \frac{L}{l}}.
\label{eq38}
\end{equation}

Our results presented in Figs.~\ref{figureWL1} and \ref{figureWL2} differ notably from the results of Ref.~\cite{araki2014weak}. Most importantly, they report a plateau in the normalized conductance $r$ at $x_\alpha\sim 0.4$. We argue that this plateau is not of physical origin, but rather an artifact of the incorrect $Q$ expansion of the Cooperons.  
\section{Application: Persistent spin helix \label{sec6}}
In addition to the Rashba case (Sec~\ref{sec3}), another high-symmetry scenario where it is possible to analytically solve the linearized Eilenberger equation is the so-called pure gauge  case
\begin{equation}
\mathcal{A}_i=m \alpha_i\mathbf{h}\cdot \boldsymbol{\sigma},
\label{eq39}
\end{equation}
where $\mathbf{h}=(h_x,h_y,h_z)$ and $\boldsymbol{\sigma}=(\sigma_x,\sigma_y,\sigma_z)$. Namely, this kind of SOC can be removed from the Eilenberger equation, or gauged-out, by a local unitary transformation of the form $U=e^{i\mathcal{A}_ir_i}$, where $\mathbf{r}=(x,y).$ \cite{tokatly2010duality, tokatly2010gauge, bernevig2006exact, koralek2009emergence}.

One of the most interesting consequences of the pure gauge SOC is the absence of spin relaxation for certain spin directions\cite{schliemann2003nonballistic}, and formation of stable spatially inhomogeneous spin structures - the so called persistent spin helices\cite{bernevig2006exact,koralek2009emergence}. 
There are two examples of the pure gauge case that are widely studied. The first one is a Dresselhaus model for the quantum wells of GaAs grown along the (110)-axis, where $\mathcal{A}_x=m\alpha \sigma_z$ and $\mathcal{A}_y=0$. The second example is the compensated Rashba+Dresselhaus model, corresponding to GaAs structures grown along the (001) axis, where $\mathcal{A}_x=\mathcal{A}_y=m\alpha (\sigma_x-\sigma_y)$. 

Without loss of generality, we  fix $\mathbf{h}\cdot \boldsymbol{\sigma}=\sigma_z$ and  determine what kinds of spatial spin structures form upon spin injection in a setup similar to Fig.~\ref{schematic}. Because there is no  spin relaxation for certain spin directions in the pure gauge case, the injected spin grows without bound in the sample. To remedy this, we will modify the setup presented in Fig.~\ref{schematic} by introducing an additional ferromagnetic electrode, which has the same magnetization and orientation as the fist one, and at the distance $L$ from it. The first electrode then serves as a source of spin, while the other one will be a spin sink. The Eilenberger equation for the pure gauge SOC in real space is then
\begin{multline}
\partial_i n_i v_F f-iv_Fn_i [m \alpha_i\sigma_z,f]\\=-\frac{1}{\tau}(f-\langle f \rangle)+G [\delta(x)+\delta(x-L)],
\label{eq40}
\end{multline}
where the two terms proportional to the $\delta$-function describe two ferromagnetic electrodes at positions $x=0$ and $x=L$, and the source term $G$ has the same meaning as in Sec.~\ref{sec4}.

We gauge-out the SOC by the unitary transformation
\begin{equation}
f=U \tilde{f} U^\dagger, \qquad G=U \tilde{G} U^\dagger, \qquad U=e^{i m r_i\alpha_i\sigma_z}.
\label{eq41}
\end{equation}
 We can now rewrite Eq.~\eqref{eq40} as
\begin{equation}
v_F n_i\partial_i \tilde{f}=-\frac{1}{\tau}(\tilde{f}-\langle\tilde{f}\rangle)+\tilde{G}[\delta(x)+\delta(x-L)].
\label{eq42}
\end{equation}
To solve Eq.~\eqref{eq42}, we transform it to the Fourier space
\begin{equation}
i Q_i n_i v_F \tilde{f}(Q)=-\frac{1}{\tau}[\tilde{f}(Q)-\langle\tilde{f}(Q)\rangle]+\tilde{G}_0+\tilde{G}_Le^{i Q_x L}.
\label{eq43}
\end{equation}
Here, we introduced $\tilde{G}_\lambda=G_z \sigma_z \delta(Q_y l)+\exp(-i \lambda \alpha_x m \sigma_z)\mathbf{G}_\bot\cdot\boldsymbol{\sigma}_\bot \exp(i\lambda \alpha_x m \sigma_z) \sum_{\pm} (1\pm \sigma_z)$ $\delta(Q_yl\pm 2m\alpha_yl)$, with $\mathbf{G}_\bot=(G_x,G_y)$ and $\boldsymbol{\sigma}_\bot=(\sigma_x,\sigma_y)$, and  $\lambda=0,L$ are the positions of the two electrodes. The solution is 
\begin{equation}
N_F\langle\tilde{f}\rangle(Q)=\tilde{S}(Q)=\frac{N_F}{\sqrt{1+Q^2l^2}-1}\bigg[\tilde{G}_0+e^{iQ_xL}\tilde{G}_L\bigg]\tau,
\label{eq44}
\end{equation}
where we introduced the spin density $S=N_F\langle f \rangle$. Transforming back to real space, we obtain
\begin{multline}
\tilde{S}(x,y)=-N_F\sum_{\lambda=0,L}\bigg[\mathcal{W}(x-\lambda,0) G_z\tau\sigma_z+\\ \mathcal{W} (x-\lambda,\alpha_y)e^{-i \alpha_x m (\lambda+y)\sigma_z}\mathbf{G}_\bot\cdot \boldsymbol{\sigma}_\bot\tau e^{i \alpha_x m (\lambda+y)\sigma_z}\bigg],
\label{eq45}
\end{multline}
where we defined the function
\begin{equation}
\mathcal{W}(x,\alpha_y)=\int \frac{d Q_x}{2\pi} e^{-i Q_x x}\frac{1}{\sqrt{1+Q_x^2l^2+4m^2\alpha_y^2l^2}-1}.
\end{equation}
The function $\mathcal{W}$ can be expressed in terms of known special functions if $\alpha_y=0$, namely,
\begin{equation}
\mathcal{W}(x,0)=-|x|/(2l)-\frac{1}{4\pi}G_{0,1}^{2,1}\left(
\begin{array}{c}
\frac{3}{2}\\
0,0,\frac{1}{2}
\end{array}\middle\vert
\frac{x^2l^2}{4}
\right),
\label{eq47}
\end{equation}
where $G_{0,1}^{2,1}(...)$ is one of the so-called Meijer G-functions \cite{bateman1953higher}. It is instructive to look at the behavior of the function $\mathcal{W}(x,\alpha_y)$ for $x\gg l$, where we may approximate 
 \begin{equation}
\mathcal{W}(x,0)\approx -\frac{|x|}{l}, \quad \mathcal{W}(x,\alpha_y)\approx \frac{e^{-2 m \alpha_y \frac{|x|}{l}}}{2 m \alpha_y l}.
\label{eq48}
\end{equation}

The transformed spin $\tilde{S}$ depends on both $x-$ and $y-$ coordinates, but the physical spin $S$ depends only on the $x$-coordinate  $S(x)=U \tilde{S}(x,y) U ^\dagger$. This is as expected, since the injection setup is homogeneous in the $y$-direction. The final expression for the physical spin is
 \begin{align}
 &S_z(x)=\sum_{\lambda=0,L}\mathcal{W}(x-\lambda,0)N_F G_z \tau, \nonumber \\
 &\mathbf{S}_\bot(x)=\sum_{\lambda=0,L}\mathcal{W}(x-\lambda,\alpha_y)
 \mathcal{R}[2m\alpha_x(x-\lambda)]N_F\mathbf{G}_\bot\tau.
 \label{eq49}
 \end{align}
 Here $\mathcal{R}(\theta)=\begin{bmatrix}\cos\theta & \sin \theta \\ -\sin \theta & \cos \theta \end{bmatrix}$ is the rotation matrix for an angle $\theta$. Note that this result could also be obtained in a more direct but less elegant way, without exploiting the gauge symmetry, by directly solving Eq.~\eqref{eq6}.

Let us discuss the results of Eq.~\eqref{eq49} by considering two scenarios: injection of the $z$-component of the spin $(G_z\neq 0, G_x=G_y=0)$, which is colinear with the gauge SOC potential, and injection of the $x$-component $(G_x\neq 0, G_y=G_z=0)$, which is perpendicular to the SOC potential.   

\begin{figure}[h!]
	\includegraphics[width=0.3\textwidth]{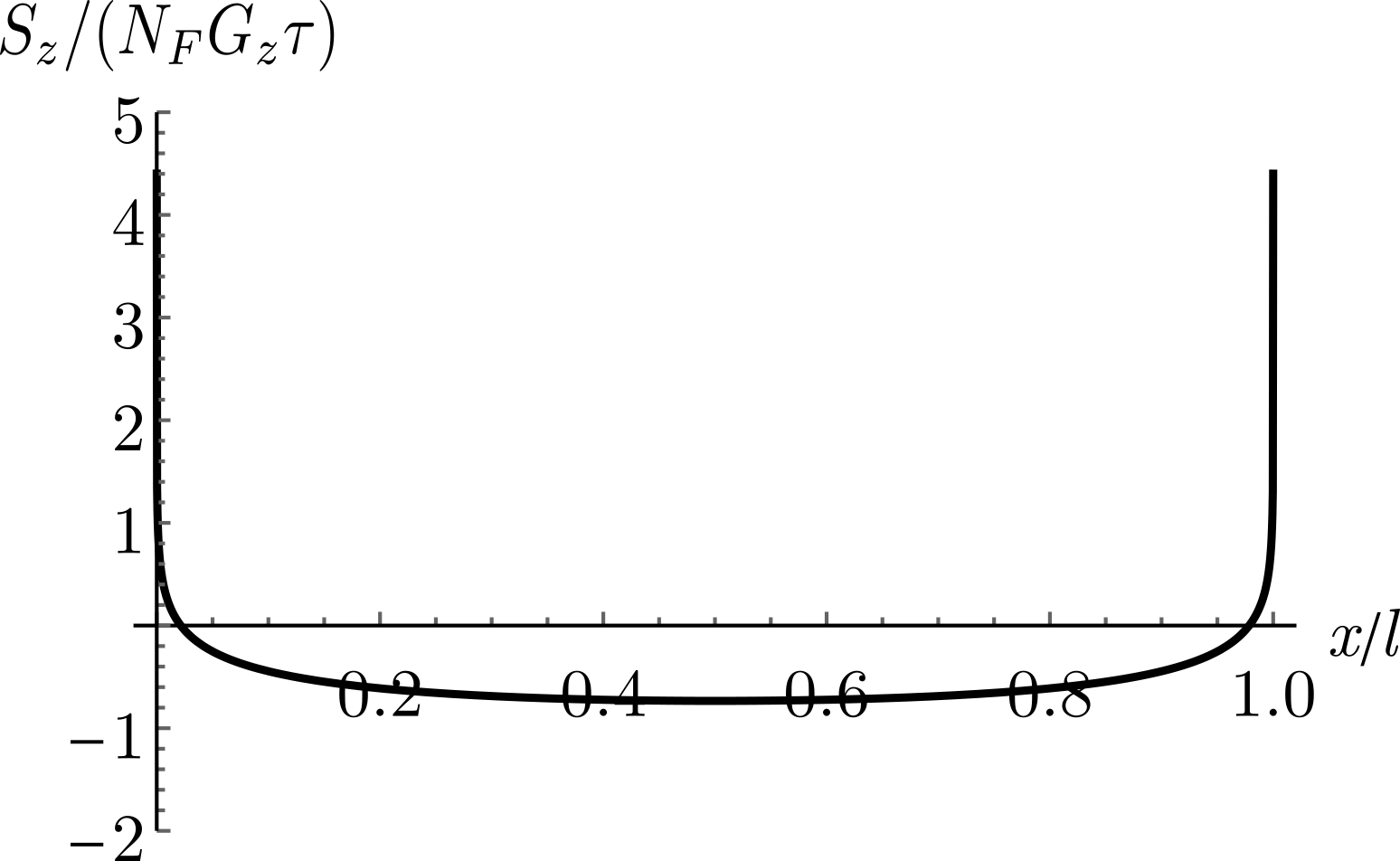}
	\caption{\label{plotgauge1} Density of the spin component $S_z$. Two ferromagnetic electrodes are placed at the positions $x$=0 and $x=l$.}
\end{figure}

\paragraph{Injection of spin polarized in $z$-direction}
$S_z$ spin component does not "feel" the gauge SOC. As a consequence, it forms an uniform spatial structure away from the electrodes, as shown in Fig.~\ref{plotgauge1}

\begin{figure*}[t!]
	\includegraphics[width=\textwidth]{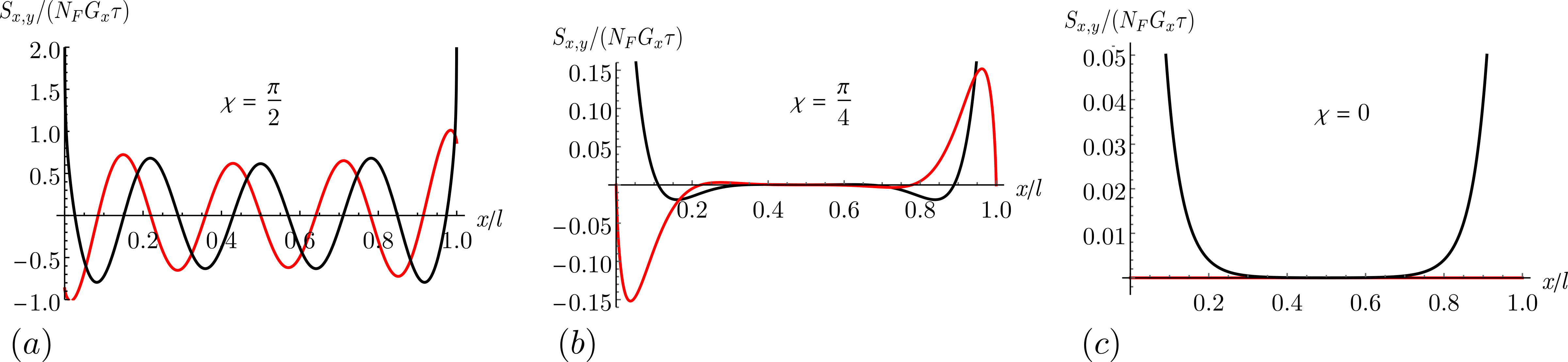}
	\caption{\label{plotgauge} Densities of the spin components $S_x$ (black curve) and $S_y$(red curve). Two ferromagnetic electrodes are placed at the positions $x$=0 and $x=l$. The magnitude of spin orbit coupling is fixed to $\sqrt{\alpha_x^2+\alpha_y^2}p_F\tau=10$ in all plots. We consider different angles $\chi$ between the injection axis and the spin-orbit field : $(a)$ $\chi=\pi/2$, $(b)$ $\chi=\pi/4$, $(c)$ $\chi=0$.}
\end{figure*}

\paragraph{Injection of spin polarized in $x$-direction}
The behavior of coupled spin components $S_x$ and $S_y$ is determined by the angle $\chi$ between the injector ($y$-axis) and the vector determined by the SOC potential ($\alpha_x,\alpha_y$): $\tan \chi=\alpha_x/\alpha_y$. As seen from the second line in Eq.~\eqref{eq49}, $\alpha_x$ and $\alpha_y$ play distinctly different roles in determining the spin densities $S_y$ and $S_x$. Namely, $\alpha_x$ only contributes to the angle of rotation in the matrix $\mathcal{R}$, and therefore introduces spatial oscillation with a period $2m\alpha_xl$. On the other hand, $\alpha_y$ enters in the function $\mathcal{W}$, which decays on the scale of $\sim m\alpha_y l$, and therefore this term is responsible  for the spin relaxation and decay of the spin density. For the case $\chi=\pi/2$, there is no spin relaxation meaning that $S_x$ and $S_y$ form a modulated spatial spin structure, better known as the persistent spin helix. On the other hand, for the case $\chi=0$, the spin component $S_x$ rapidly decays and $S_y$ is not induced. Fig.~\ref{plotgauge} illustrates the behavior of spin components $S_x$ and $S_y$ for several values of the angle $\chi$.   

\section{Conclusion}

Eilenberger equation is a well known tool used in the study of superconductivity In this work we demonstrate that it can be effectively used in the normal state as well, where it provides an intuitive tool to study spin transport and weak localization at any degree of disorder. In Sec.~\ref{sec2}, we formulated  the linearized Eilenberger equation for any linear-in-momentum SOC using the covariant SU(2) formalism [Eq.~\eqref{eq2}], and provided a generic solution in terms of Fermi surface averages [Eq.~\eqref{eq6}]. For the specific case of Rashba SOC, this yields a relatively simple closed-form solution, which we elaborate in Sec.~\ref{sec3}. We used this Rashba solution to address three unrelated  problems.

 Firstly, we studied spin injection problem by a ferromagnetic electrode. We calculated the spatial distribution of spin and charge density upon spin injection at arbitrary disorder. In the case when the injected spin direction is colinear with the electrode, we demonstrate a "spin battery" effect in Sec.~\ref{subsec4B}.  
 
 Secondly, we demonstrated the power of our approach to study magnetoelectric phenomena in superconductors on the example of the superconducting Edelstein effect. Starting from our general solution (Sec.~\ref{sec3}), we recover this effect in just a few lines of calculation. Moreover, we generalize previously known results in the ballistic \cite{edelstein1995magnetoelectric} and the diffusive limit \cite{edelstein2005magnetoelectric}.
 
 Thirdly, we addressed the problem of weak localization in the Rashba conductor beyond the diffusive limit (Sec.~\ref{sec5}), and corrected previous works on this topic. More importantly, we demonstrated a way to avoid cumbersome diagrammatic calculations and obtain the results in a more transparent manner. This approach could be useful to describe systems with other kinds of SOC, for instance Rashba+Dresselhaus model, where W(A)L is lately intensively studied both theoretically and experimentally due to a potential to realize persistent spin helix structures \cite{miller2003gate}.  
 
 Furthermore, we solved the Eilenberger equation for the so-called pure gauge case in Sec.~\ref{sec6}. We studied the formation of spin textures upon local spin injection, and demonstrated that they greatly depend on the relative orientation between the injector and the effective SOC vector potential. 
 
 Our work establishes a direct relationship between several different phenomena mediated by the SOC: spin-triplet superconductivity, spin transport and weak localization. The presented equations can be used to study all these phenomena in various systems, such as hybrid nanostructures and inhomogeneous systems, and they can be adapted to address novel materials such as transition metal dichalcogenide monolayers, different geometries and physical situations.   
 
 \section*{Acknowledgments}
We acknowledge funding from Spanish Ministerio de Ciencia,
Innovación y Universidades (MICINN) (Projects No. FIS2016-79464-P and
No. FIS2017-82804- P).  I.V.T. acknowledges support by Grupos Consolidados
UPV/EHU del Gobierno Vasco (Grant No. IT1249-19).   S. I. and F.S.B  acknowledge  funding from EUs Horizon 2020 research and innovation program under Grant
Agreement No. 800923 (SUPERTED).
  
\appendix
\section{Weight factor in weak localization \label{app1}}
\begin{figure}[h!]
	\includegraphics[width=0.4\textwidth]{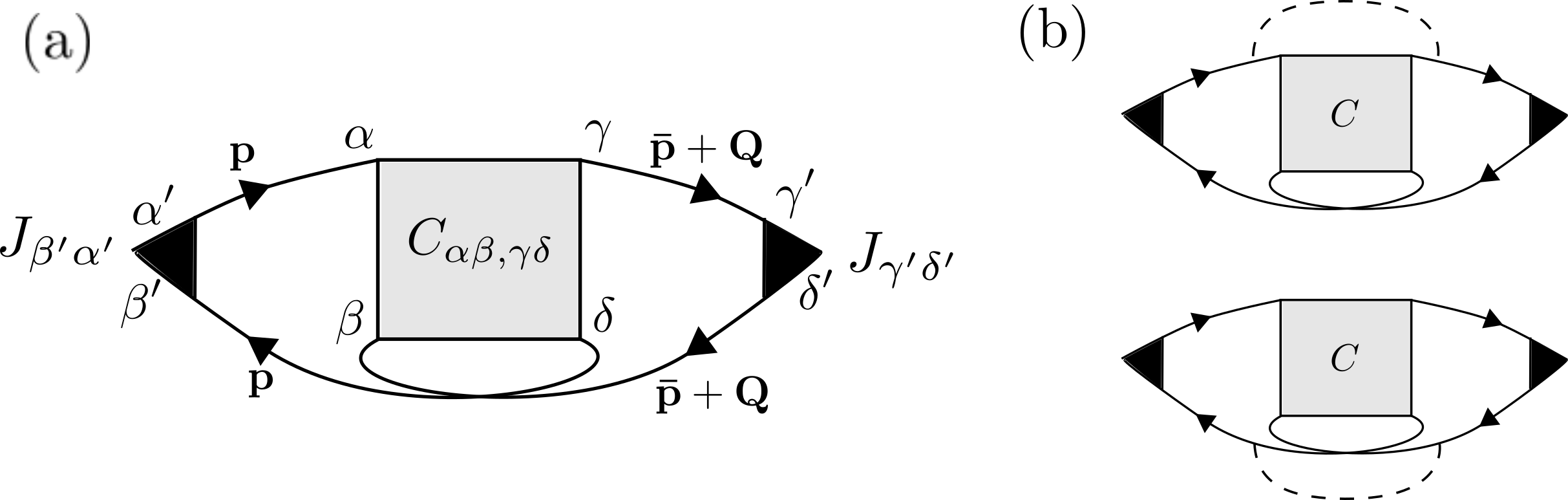}
	\caption{\label{figapp}Diagrams for the WL correction to the conductance. Solid arrows represent Green's functions while the dashed lines represent disorder. The upper (lower) branch of the diagrams corresponds to retarded (advanced) Green's functions. Vertices correspond to the renormalized current operator. (a) Bare Hikami box. (b) Dressed Hikami boxes. Greek indices describe the spin degree of freedom.  }
\end{figure}

In this Appendix, we prove Eqs.\eqref{eq32} and \eqref{eq33}  from the main text using the diagrammatic perturbation theory. First, we define the advanced and retarded Green's functions which will be employed in the diagrams 
\begin{equation}
G^{R,A}_\mathbf{p}=\bigg(\xi_\mathbf{p}+m \alpha p_F \sigma_y\cos \theta-m \alpha p_F \sigma_x\sin \theta\pm \frac{i}{2\tau} \bigg)^{-1}.
\label{eqa1}
\end{equation}
Next, we introduce the renormalized current operator \cite{guerci2016spin} as 
\begin{equation}
J_{x\mathbf{q}}=v_F\cos\phi.
\label{eqa2}
\end{equation}
Diagrammatic representation of the weak localization correction to the conductance in terms of maximally crossed diagrams - Cooperons $C$, is given in Fig.~\ref{figapp}.

We distinguish two contributions to the WL conductance
\begin{equation}
\delta\sigma=\delta\sigma^{(a)}+\delta\sigma^{(b)},
\label{eqa3}
\end{equation}
where $\delta\sigma^{(a)}$ comes from the so-called bare Hikami box\cite{akkermans2007mesoscopic} [Fig.~\ref{figapp}(a)], and $\delta\sigma^{(b)}$ comes from the dressed Hikami boxes [Fig.\ref{figapp}~(b)]. Explicitly evaluating diagrams in Fig.~\ref{figapp} (a) yields
\begin{multline}
\delta \sigma^{(a)}=\frac{e^2}{2\pi}\int \frac{d^2\mathbf{p}}{(2\pi)^2}\int \frac{d^2\mathbf{Q}}{(2\pi)^2}[G^R_\mathbf{p}]_{\alpha'\alpha}C_{\alpha\beta,\gamma\delta}(\mathbf{Q})\\\times[G_{\mathbf{\bar{p}+Q}}^R]_{\gamma\gamma'}[J_{x\mathbf{\bar{p}+Q}}]_{\gamma'\delta'}[G_{\mathbf{\bar{p}+Q}}^A]_{\delta'\beta'}[G_{\mathbf{p}}^A]_{\delta\beta}[J_{x\mathbf{p}}]_{\beta'\alpha'},
\label{eqa4}
\end{multline}
where the summation over repeated indices is assumed. The Cooperons $C_{\alpha\beta,\gamma\delta}$ that enters Eq.~\eqref{eqa4}  needs to be transformed from the basis of spin indices to the singlet-triplet basis, which is the basis used in the main text. This is achieved by the following transformation \cite{akkermans2007mesoscopic,mccann2012z}
\begin{equation}
C_{ss'}=\frac{1}{2} [\sigma_y\sigma_s]_{\alpha\beta}C_{\alpha\beta,\alpha'\beta'}.
[\sigma_{s'}\sigma_y]_{\beta'\alpha'}.
\label{eqa5}
\end{equation}
Applying the transformation to Eq.~\eqref{eqa4}, we obtain
\begin{equation}
\delta\sigma^{(a)}=\frac{e^2}{2\pi}\int \frac{d^2\mathbf{Q}}{(2\pi)^2}\text{Tr}[C(\mathbf{Q}) W^{(a)}],
\label{eqa6}
\end{equation}
where $W$ is the weight factor matrix given as
\begin{multline}
W_{ss'}^{(a)}=\frac{1}{2} \int \frac{d^2\mathbf{p}}{(2\pi)^2}\\\text{Tr}\bigg[\sigma_y\sigma_{s'} G^A_\mathbf{p}J_{x\mathbf{p}}G^R_{\mathbf{p}} \bigg( G^R_\mathbf{\bar{p}}J_{x\mathbf{\bar{p}}}G^A_{\mathbf{\bar{p}}} \sigma_y\sigma_{s}\bigg)^T\bigg].
\label{eqa7}
\end{multline}
Note that here we neglected the weak $Q$-dependence of the weight factor, which is justified since the dominant contribution of the Cooperons comes from small $Q$. 

Similarly $\delta \sigma^{(b)}$ is obtained using the expression \eqref{eqa6} with the weight factor substituted by $W^{(b)}$, given as
\begin{multline}
W_{ss'}^{(b)}(\mathbf{Q})=\frac{1}{2} \int \frac{d^2\mathbf{p}}{(2\pi)^2} \int \frac{d^2\mathbf{p'}}{(2\pi)^2} \text{Tr}\bigg[\\
\sigma_y\sigma_{s'} G^A_\mathbf{p}J_{x\mathbf{p}}G^R_{\mathbf{p}}G^R_{\mathbf{p'}} \bigg(G^R_\mathbf{\bar{p}} G^R_\mathbf{\bar{p}'}J_{x\mathbf{\bar{p}'}}G^A_{\mathbf{\bar{p}'}} \sigma_y\sigma_{s}\bigg)^T \\
+\sigma_y\sigma_{s'} G^A_\mathbf{p'}G^A_\mathbf{p}J_{x\mathbf{p}}G^R_{\mathbf{p}} \bigg( G^R_\mathbf{\bar{p'}}J_{x\mathbf{\bar{p'}}}G^A_{\mathbf{\bar{p'}}}G^A_{\mathbf{\bar{p}}} \sigma_y\sigma_{s}\bigg)^T\bigg].
\label{eqa8}
\end{multline}
Here, the first and second line come from the two different types of dressed Hikami boxes, represented in the upper and lower panel of Fig.~\ref{figapp} (b), respectively. They give equal contributions after integration.

Finally, after performing the momentum integration in Eqs.~\eqref{eqa7} and \eqref{eqa8}, we arrive at Eq.~\eqref{eq33} in the main text. Note that $W_0$ and $W_y$, as well as first term in $W_x$ and $W_z$, come from bare Hikami boxes, while the remaining terms come from dressed Hikami boxes.  
\section{Coefficients in weak localization \label{app2}}
In this Appendix we write the expansion coefficients for quantities $a,b,c,d$ and $e$, introduced above Eq.~\eqref{eq35} in the main text. This is followed by the definition of the relaxation lengths $\lambda_i$ and the coefficients $A_{x,zi}$ that appear in Eq.~\eqref{eq35}.

The expansion coefficients are
\begin{widetext}
\begin{align}
&a_0=\frac{1+\frac{x_\alpha^2}{2}}{1+x_\alpha^2},\quad a_2=-\frac{2+3x_\alpha^4+x_\alpha^6}{4(1+x_\alpha^2)^3}, \quad a_4=\frac{3}{16}
\frac{2-5x_\alpha^2+15x_\alpha^4+10x_\alpha^6+5 x_\alpha^8+x_\alpha^{10}}{(1+x_\alpha^2)^5}, \quad b_2=\frac{x_\alpha^2(6+3x_\alpha^2+x_\alpha^4)}{8(1+x_\alpha^2)^3},\nonumber \\  & b_4=-\frac{x_\alpha^2(15+5x_\alpha^2+10x_\alpha^4+5x_\alpha^6+x_\alpha^8)}{8(1+x_\alpha^2)^5}, \quad c_0=\frac{1}{1+x_\alpha^2},\quad c_2=\frac{-1+3x_\alpha^2}{(1+x_\alpha^2)^3},\quad c_4=\frac{3(1-10x_\alpha^2+5x_\alpha^4)}{8(1+x_\alpha^2)^5}, \nonumber \\
&d_0=1, \quad d_2=-\frac{Q^2l^2}{2},\quad d_4=\frac{3Q^4l^4}{8}\quad, \quad  e_1=-\frac{x_\alpha}{(1+x_\alpha^2)^2},\quad e_3=-\frac{3x_\alpha (-1+x_\alpha^2)}{2(1+x_\alpha^2)^4}.
\label{eqb1}
\end{align}

Using these coefficients, we may approximate the Cooperons from Eq.~(34) of the main text as
\begin{equation}
C_x(Q)=C_y(Q)=\frac{1}{4\pi N_F\tau}\bigg[\frac{1}{1-a_0+Q^2l^2(b_2-a_2)}+\frac{1-c_0-c_2Q^2l^2}{\alpha+\beta Q^2l^2+\gamma Q^4l^4}\bigg], \, C_z(Q)=\frac{1}{2\pi N_F\tau}\frac{1-a_0-Q^2l^2(a_2+b_2)}{\alpha+\beta Q^2l^2+\gamma Q^4l^4}.
\label{eqb2}
\end{equation}
\end{widetext}
Here, $\alpha=(1-a_0)(1-c_0)$, $\beta=-(a_2+b_2)(1-c_0)-(1-a_0)c_2-d_1^2$, $\gamma=-(a_4+b_4)(1-c_0)+(a_2+b_2)c_2-(1-a_0)c_4-2d_1d_3$. After performing the partial fraction decomposition, Eq.~\eqref{eqb2} reduces to Eq.~\eqref{eq35} from the main text, where relaxation lengths $\lambda_i$ are
\begin{equation}
\lambda_{1,2}^{-2}=\frac{1}{l^2}\frac{\beta\mp\sqrt{\beta^2-4\alpha\gamma}}{2\gamma},\qquad \lambda_3^{-2}=\frac{1}{l^2}\frac{1-a_0}{b_2-a_2},
\label{eqb3}
\end{equation}
and the coefficients in the decomposition are
\begin{align}
&A_{x1,2}=\mp \frac{1-c_0+c_2\lambda_{1,2}^{-2}l^2}{2\gamma l^2(\lambda_1^{-2}-\lambda_2^{-2})}, \qquad A_{x3}=\frac{1}{2(b_2-a_2)},\nonumber \\ &A_{z1,2}=\mp \frac{1-a_0+(a_2+b_2)\lambda_{1,2}^{-2}l^2}{\gamma l^2(\lambda_1^{-2}-\lambda_2^{-2})}.
\label{eqb4}
\end{align}
\end{document}